\documentclass[printer]{aa}
\usepackage{txfonts}
\usepackage{natbib}
\bibpunct{(}{)}{;}{a}{}{,} % to follow the A&A style\usepackage{mathcomp}
\usepackage{graphicx}%,amsmath,multicol,epsfig,floatflt}
\usepackage{tabularx}

\newcommand{\gp}{giant planets }
\newcommand{\mj}{\,M$_\mathrm{Jup}~$}
\newcommand{\dg}{\,^{\circ}}

\begin{document}

\title{High resolution imaging of young M-type
  stars of the solar neighbourhood: Probing the existence of
  companions down to the mass of Jupiter.
\thanks{  Based on observations made with the
NACO at VLT UT-4 at the Paranal Observatory under
programs ID  084.C-0739, 085.C-0675(A), 087.C-0413(A) and 087.C-0450(B). 
}} 
 
%\subtitle{II. An example text with infinitesimal
%  scientific value\\
%  whose title and subtitle may also be split}

\author{
  P. Delorme \inst{1}
  \and A.M. Lagrange \inst{1}
  \and G. Chauvin \inst{1,2}
  \and M. Bonavita\inst{3}
   \and S. Lacour \inst{4}
 \and M. Bonnefoy \inst{2}
 \and   D. Ehrenreich \inst{1}
 \and  H. Beust \inst{1}
}

\offprints{P. Delorme, \email{Philippe.Delorme@obs.ujf-grenoble.fr}}

\institute{UJF-Grenoble 1 / CNRS-INSU, Institut de Plan\'etologie et d'Astrophysique de Grenoble (IPAG) UMR 5274, Grenoble, F-38041, France.
  \and Max Planck Institute for Astronomy, K\"onigstuhl 17, D-69117 Heidelberg, Germany. 
  \and Department of Astronomy and Astrophysics, University of
  Toronto, 50 St. George Street, Toronto, Ontario, Canada M5S 3H4
  \and LESIA, CNRS/UMR-8109, Observatoire de Paris, UPMC, Universit\'e Paris
Diderot, 5 place Jules Janssen, 92195 Meudon, France.
}

\abstract{
High contrast imaging is a powerful technique to search for gas \gp
 and brown dwarfs orbiting at separation larger than several AU.
 Around solar-type stars, \gp are expected to form by core accretion or
 by gravitational instability, but since core accretion
 is increasingly difficult as the primary star becomes lighter,
 gravitational instability would be the a probable formation
 scenario for yet-to-be-found distant \gp around a low-mass star. A
 systematic survey for
 such planets around M dwarfs would therefore provide a direct test of the
 efficiency of gravitational instability.}
{
We search for gas \gp orbiting around late-type
  stars and brown dwarfs of the solar neighbourhood.}
{We obtained deep high resolution images of 16 targets with the adaptive
  optic system of VLT-NACO in the $L'$ band, using direct imaging and angular
  differential imaging.  This is currently the largest and deepest
  survey for Jupiter-mass planets around M-dwarfs.
We developed and used an integrated reduction
and analysis pipeline to reduce the images and derive our 2D detection
limits for each target. The typical contrast achieved is about 9
magnitudes at 0.5$\arcsec$ and 11 magnitudes beyond 1$\arcsec$. For
each target we also determine the 
probability of detecting a planet of a given mass at a given
separation in our images.} 
{ We derived accurate detection
  probabilities for planetary companions, taking into account orbital
  projection effects, with in average more than
  50\% probability to detect a 3\mj companion at 10\,AU and a 1.5\mj
  companion at 20\,AU, bringing strong constraints on the existence of Jupiter-mass planets
  around this sample of young M-dwarfs. }

\date{}


\titlerunning{High resolution imaging of young M-type
  stars of the solar neighbourhood}

\maketitle

\section{Introduction}
 The discovery of extrasolar planets around solar analogs, initiated by \citet{Mayor.1995},
 has radically modified our understanding of planetary formation. The
 new models that strive to describe planetary formation mechanisms
 need observational constraints on planetary companions on a large
 range of parameters, such as stellar mass, metallicity, planet mass and
 orbital parameters. Notably, the most prolific methods for exoplanet
 detection, radial velocities and transits, only cover separations up
 to a few AU and barely reaches the large separation range where all
 the Solar System \gp reside. This range can be probed by direct
 high-resolution imaging. 
 Only a few  unambiguously planetary-mass companions to stars \citep{Chauvin.2004,MaroisHR8799.2008,Kalas.2008,Lagrange.2010} were discovered by
 direct imaging, but each one of these
 detections proved extremely valuable to constrain formation models. 
 Until recently, 
  core accretion (CA) was the preferred model to explain both our
  Solar System planets and the close-in giant planets found by
  radial velocity (RV) around solar-type  stars. 
The recently imaged
  \gp at large separations from their parent A-type stars 
\citep{MaroisHR8799.2008,Kalas.2008}
  challenge however this simple view as they are very difficult to form {\it in
    situ} by CA, though \citet{Rafikov.2011} shows that under specific
  condition CA can form Jupiter-mass planets as far as 40-50AU. In the
  case of HR8799, this could account for the formation {\it in situ} of the
  3 innermost planets, but \citet{Dodson-Robinson.2009} claim that
  HR8799bcd formation by CA has to happen at closer separation and be
  followed by planet scattering at the observed separation, which would not
  create a stable system. \citet{Dodson-Robinson.2009} claim that, instead, gravitational instability (GI) in
  a disc could produce the $bcd$ planets {\it in situ} while
  \citet{Kratter.2010} argue that HR8799bcd can be formed by GI, but
  only at larger
  separation and show that under certain
  conditions they can migrate to the separations where they are
  currently observed. However the innermost planet of this system,
  HR8799e \citep{Marois.2010HR8799e} appears at least as difficult to form by GI as the outermost
  planet HR8799b would be to form by CA, hinting that planetary
  formation models still need improvement to be able to account for
  the first multi-exoplanetary system directly imaged and badly
  need further observational constraints, especially in different mass
  range, before drawing any strong conclusion
  on the respective part they actually play in planetary formation. 

Planetary mass objects could also form in the same way as
  stars, with  \citet{Bate.2009} showing that
  a few objects as light as $\sim$5\mj can form by core fragmentation,
  possibly ending in very high mass-ratio binary system, very look-alike to
  a planetary system. Such planetary-like binary system could also be
  formed by disc
  instabilities in a very massive primordial disc
  \citep{Stamatellos.2011}, which could produce systems quite similar
  to 2M1207AB
  \citep{Chauvin.2004}, together with numerous brown dwarfs and very
  low-mass stars.

   The dependency of planetary formation modes to the stellar mass
   is an open field of investigation, with some observational clues
   that planetary masses, as well as disc masses, scale with stellar
   masses \citep[e.g.][]{Forveille.2011,Scholz.2006}. Debris and
   transitional discs 
   tracing planetary formation have been detected even around M
   dwarfs and brown dwarfs, with
   frequencies similar \citep{Plavchan.2009} or even larger
   \citep{Currie.2011} than around solar-type stars.

\citet[][]{Boley.2009} identifies some indications of
  bi-modal \gp formation, but many more detections are needed to confirm this
  scenario. 
In that frame, looking for gas giant planets around M-type stars
offers a distinct opportunity to test planet formation
mechanisms, as models predict that CA is very inefficient
\citep[][KK08]{Kennedy.2008}. KK08 predict a rate of 1$\%$ for CA formed
\gp around M stars if the rate is of 6$\%$ for G type stars) or ``all
but impossible" (DR09) in producing \gp around those stars. The main
reason being that there is not enough time to form cores with 5-10M$_{\oplus}$
before the gas in the protoplanetary disk is dissipated.
% We note that according to KK08, ``failed cores" with masses of a few
% Earth-masses can form on longer timescales and be still present
% around those stars.
 On the contrary, GI can be quite efficient in forming \gp around M
 stars, at separations larger than typically 20\,AU with \citet{Boss.2011} claiming that ``even late M dwarf stars might be able to
        form gas giants on wide orbits". This hints at the possibility
        that there are more \gp at large
   separations around late-type
   stars than there are close-in ones. 
% This theoretical ground is supported by the empirical scaling
%  down of the \gps observed around HR8799: such $\sim$ 10MJup objects
%  translate into a few Jupiter mass \gp around M dwarfs, which would be
%  detectable with NACO. 

From an observational point of view, most of the planets found so far
 around a yet small number of M-type stars  are rather low-mass ($\leq$
 0.1 \mj) ones, and in close orbits.
 The monitoring of
 M-type stars started recently and these active stars are complicated
 targets for RV searches; hence, the domain $\geq$ 3\,AU is largely
 unexplored via RV technics. However the trend that massive \gp are
 less common around M stars than around more massive stars at short
 separations seems statistically robust \citep[e.g.][; Bonfils et
   al. 2011, submitted]{Forveille.2011}. The picture at the larger
 separations ($\geq$ 10\,AU) is less clear cut. First, RV survey show
 that planetary frequency around M dwarfs increases as separation
 increases (Bonfils et al. 2011, submitted). Secondly this trend is
 supported by evidence from micro-lensing surveys, with
 \citet{Gould.2010} showing that the frequency of Jupiter (respectively
 Saturn) mass planets at $\sim$10\,AU is about 10\% (respectively
 20\%), and correlates well with the extrapolation of RV surveys to
 large separations.

 This separation range 
 started being explored with direct Adaptive Optic (AO) imaging in the recent years, on a
 small number of late type objects \citep[17 out of 118 in the recent compilation of][]{Nielsen.2010}, and with limited sensitivities (typically $>$4
 \mj). Others surveys targeted at young late M stars and brown dwarfs were very
 sensitive to unequal mass binary systems \citep[$q \sim 0.2$;][]{Ahmic.2007,Biller.2011} but did not
 reach mass ratio typical of planetary systems ($q < 0.01$), and
 therefore did not probe the supposedly more numerous light \gp
 ($\leq$1\mj) that
 are supposed to be more frequent at large separation around M dwarfs. 

The
 main limitation of high resolution imaging of M-dwarfs is their
 intrinsic faintness which makes difficult to achieve good AO correction from the ground. Spatial observations
 \citep[e.g.][]{Bouy.2003} and Laser Guide Star observations
 \citep[e.g.][]{Biller.2011} do mitigate this issue but make
 high-resolution deep
 observations of  M dwarf more challenging than similar observations
 around solar-type stars. However, since most of  M dwarfs flux is
 emitted in the infrared, the capabilities of VLT-NAOS
 adaptive optics system with Infrared Wave-Front Sensor (IRWFS) significantly
 broaden the sample of M dwarfs which can be used as a good quality
 natural guide star, thus enabling their observations at high Strehl
 ratio from the ground. Given the very red Spectral Energy Distribution (SED) of Jupiter mass planets, typically
 $Ks-L'$=3.3 at 3\mj and $Ks-L'>$6 at 1\mj at 30Myr \citep[according
   to BT-SETTL 2010 models,][]{Allard.2010}, observing in the
 $L'$ band offers a greatly enhanced sensitivity to the lowest mass
 planets.
 The additional capability of NACO to observe
 in the $L'$ range therefore makes NACO at VLT a unique instrument to probe for
 planets around M dwarfs. The high spatial density of M-type stars in the solar
 neighbourhood makes it easier to build a good sample of young and
 nearby low-mass stars than it is for solar-type stars. Their youth
 ensures a more favourable contrast for planetary companions and their
 intrinsically low flux enable to probe much lower companion masses around M
 dwarfs than around any other stars for a given contrast.  
We therefore started a direct imaging survey in $L'$ with NACO to look for \gp around
young nearby M-dwarfs in stellar associations, and probe
 separations down to a few AU and masses down to $\sim$0.5 \mj. We describe
 our sample and observations in section 2, and detail our data analysis methods
 in section 3. Our results are presented in section 4.

\section{The stars}
\subsection{Sample selection}

  We built a sample large enough to be statistically
  significant of young late-type stars 
  close and young enough to ensure optimal detection and separation
  limits.  We selected all late-type stars younger than 50 Myr
  (for  early M stars) and 100 Myr (for $\geq$ M4V stars), closer than
  45 pc, among the members of nearby
  young associations, identified notably by \citet{Torres.2008,Lepine.2009,Kiss.2011}. For the latest type
  targets, we furthermore kept 
  only those brighter than Ks = 12, to have a good AO correction.  We
  removed targets members of known low contrast, seeing-resolved, close binary systems with 
  separations between 0.5$\arcsec$ and  1.5$\arcsec$ that could confuse the
  wave front analysis and lower the
  quality of the AO correction. We ended up with a sample of 52 targets (46
  M0-M5 and 6 M8+). We present here observations of 16 stars
  from this sample,
  detailed on  table \ref{targets}, which were observed
  during our first observing nights for this program.
% Given their K mag. and predicted Ks-L' (Lyon's
%   models), their age and distance, and assuming a $\Delta$ L' of 7.5
%   at 0.3" (Lagrange et al, 2010), and $\Delta$ L' of 10. at $\geq$
%   0.5" (Fig. 2), we can detect, if present, {\bf all GPs more massive
%     than $\sim$3 MJup at sep. $\simeq$ 0.3-0.5"}, and {\bf all planets down
%     to $\sim$1 MJup at sep. $\geq$ 0.5" (ie 5-20\,AU)}, see Fig.1.

\subsection{Observations}
  \label{observations}  

Since
 low-mass stars are fainter in the visible and brighter in the
 infrared, we used NACO with its Infrared Wave Front Sensor (IWFS) to
 achieve good adaptive optic correction on these late-type targets. To
 lessen the contrast between stars and companions, we observed at
 3.8$\mu m$, in the $L'$ band, where planets are brighter than in the
 near infrared. Depending on the brightness and hour angle of each
 targets at the time of the observations, we conducted either
 classical imaging (9 targets) and pupil tracking observations (7 targets).

   We observed 11 targets in $L'$ with NACO at VLT during the run
   084.C-0739, from December 25th to December 29th, 2009 under average
   conditions, with seeing varying from 0.6 to 1.2$\arcsec$, with
   a few excursions down to 0.5$\arcsec$ and up to 1.5$\arcsec$. During run
   085.C-0675(A), on July 28th, 2010, 2 additional
   objects were observed under moderate seeing conditions ($\geq 1
   \arcsec$), and variable absorption. Three new objects
   were observed during run 087.C-0413(A) under
   excellent conditions, with seeing varying in the 0.4-0.8$\arcsec$
   range. Finally we obtained new $L'$-band observation of one of our
   target with NACO at VLT during run 087.C-0450(B) on September 1st,
   2011, with seeing conditions of $\sim$0.8$\arcsec$. 

   All objects presented here were observed in $L'$ band with NACO,
   using the L27 camera, giving a 0.027$\arcsec$ per pixel sampling,
   and the ``Uncorrelated and high well depth" detector mode. As
   detailed in table 
   \ref{targets}, we recorded data cubes of 100 to 400 very
short exposures (0.075s to 0.4s), at the same position, totalling
typically 30-60s time on target per cube. Several cubes were obtained,
with a typical total exposure time
of $\sim$30 minutes, achieved
following a 4-position dither pattern to ensure correct background
subtraction. The AO loop was closed
on these relatively faint and red targets
using the infrared-wave front sensor with the ``$JHK$" dichroic.

\begin{table*}
 \centering
 \caption{Targets information. Pupil tracking indicates observations
   were carried out in pupil tracking and with a field rotation
   enabling ADI reduction procedures.\label{targets}}
\begin{tabular}{c|c c|c|c|c|c|c} \hline \hline
Target Name &\multicolumn{2}{c|}{Coordinates} &Spectral& Membership & Date Obs.& Exposure Time  & Comments\\ 
            &  RA  &  Dec      &     Type         &        &            & (Seconds) &  \\ \hline
%  TYC1186-706-1 &00 23 34.66 +20 14 28.7 &  F7.5    &    Bpic   & 29/12/2009 & 80x150x0.2      &  low contrast Binary  \\ \hline %http://iopscience.iop.org/1538-3881/137/3/3632/fulltext
GSC08056  &02 36 51.5 & -52 03 04.4 &  M2     &  Tuc-Hor   &   26/12/2009         & 91$\times$150$\times$0.2     & Pupil Tracking       \\ \hline
GJ3305            &04 37 37.5 & -02 29 28.2              &    M0
&$\beta$pic           & 26/12/2009      &64$\times$300$\times$0.2   & Binary; Pupil Tracking    \\ \hline %http://cdsbib.u-strasbg.fr/cgi-bin/cdsbib?2006AJ....132..866R
GJ3305            &04 37 37.5 & -02 29 28.2              &    M0
&$\beta$pic           & 01/09/2011      &250$\times$150$\times$0.2   & Binary; Pupil Tracking    \\ \hline% HIP21547 &     04 37 36.13 -02 28 24.8    &   Not VLM:drop?     & beta pic      &  28/12/2009         &  48x100x0.2      &    \\ \hline%probable companion to GJ3305:
2MASS J0443     &04 43 37.6 & +00 02 05.2        &  M9    &  (1)  &     27/12/2009       &   50$\times$300$\times$0.2    &  Pupil Tracking   \\ \hline %http://iopscience.iop.org/0004-637X/703/1/399/pdf/0004-637X_703_1_399.pdf   http://iopscience.iop.org/0004-637X/705/2/1416/pdf/0004-637X_705_2_1416.pdf
V1005 Ori  &04 59 34.8 & +01 47 00.7      &  M0    &  $\beta$pic       &   25/12/2009         &   16$\times$150$\times$0.2    &   -  \\ \hline
CD-571054  &05 00 47.1	& -57 15 25.5        &  M0V    &     $\beta$pic   & 29/12/2009           & 64$\times$150$\times$0.2    &    - \\ \hline
BD-21 1074A&05 06 49.6 & -21 35 06.0       &  M2V    & $\beta$pic       & 28/12/2009           &32$\times$120$\times$0.25      & -   \\ \hline
% HIP26966  & 	05 43 21.67 -18 33 26.9        &   A0 Not VLM:drop?    &   Tuc-Hor       &   27/12/2009       &16x150x0.2      &    \\ \hline
% HIP30030 V1358 Ori  & 	06 19 08.06 -03 26 20.4         &  F9V  Not VLM:drop?     & columba      &      28/12/2009       &  16x100x0.2     &    \\ \hline %http://cdsbib.u-strasbg.fr/cgi-bin/cdsbib?2001MNRAS.328...45M
% HIP30034  & 06 19 12.9 -58 03 15.5        &  K1    & Tuc-Hor      &  25/12/2009           &  32$\times$100$\times$ 0.2   &    \\ \hline
% HIP32435  &   	06 46 13.54 -83 59 29.5      &  F5V  Not VLM:drop? &   Tuc-Hor    &   28/12/2009         &  32x100x0.2    & Companion with contrast 10   \\ \hline
 %HD67945 & 	08 09 38.60 -20 13 49.8        & F0   Not VLM:drop?    &   Argus association      &    25/12/2009      &  48x70x0.2     &ADI    \\ \hline
BD01 2447     &10 28 55.6 & +00 50 27.6    &   M2.5V   &  AB Dor    & 02/05/2011&  48$\times$200$\times$0.2     &   Pupil Tracking \\ \hline
 2M1139      &11 39 51.1 & -31 59 21.5        &   M8   & TWA      & 26/12/2009  &  36$\times$600$\times$0.3     & -  \\ \hline
2MASS1207&12 07 33.5 & -39 32 53.9 & M8&  	TWA       & 26/12/2009       & 141$\times$300$\times$0.2       &  - \\ \hline
TWA25     &12 15 30.7 & -39 48 42.6    &   M1V   &  TWA    & 03/05/2011&    24$\times$100$\times$0.4   &  -  \\ \hline
TYC7443-1102-1  &19 56 04.4 & -32 07 37.7        &   M0V   &  $\beta$pic     & 03/05/2011&   87$\times$200$\times$0.2    & Pupil Tracking   \\ \hline
HIP102409/AUmic  &20 45 09.5 & -31 20 27.2&   M1V   &  $\beta$pic    &
28/07/2010 & 94$\times$100$\times$0.2      &   Pupil Tracking \\ \hline
 WWPsA           &22 44 58.0 & -33 15 01.7       & M1     &  $\beta$pic      &  27/12/2009        &  32$\times$150$\times$0.2     &   - \\ \hline
 TXPsA&22 45 00.0 & -33 15 25.8           &  M5V       &  $\beta$pic    &   26/12/2009      & 32$\times$150$\times$ 0.2  &  -\\ \hline
HIP114046  &23 05 52.0 & -35 51 11.1&   M2V   &  (2)  &
28/07/2010 & 92$\times$100$\times$0.2      &   Pupil Tracking \\ \hline
BD-13 64 24  &23 32 30.9 & -12 15 51.4       &   M0V   &  $\beta$pic     & 27/12/2009& 32$\times$150$\times$0.2      &  -  \\ \hline
\end{tabular}
\tablebib{(1) Presence of lithium according to \citet{Reiners.2009} and low
  gravity \citep{Cruz.2007} ensures this object is younger than
  120Myrs (see section 2.3.2).
(2) Age determination is presented in section 2.3.3. For mass limit determination purposes we  assumed an age range of 100-2000 Myr.}
\end{table*}

\subsection{Estimation of the age of the stars in the sample}
To convert the observed companions fluxes into masses or to estimate
the detection probability in term of masses, one needs to use
brightness-mass relations, which are strongly dependent on the age of
the system.

  \subsubsection{Targets belonging to young moving groups}
In this case the age determination is relatively straightforward since
we can assume that a star has the age of its parent moving
group. Following \citet{Torres.2008} and references therein we
assumed an age of 8 Myr for members of TW Hydrae, 12 Myr for
$\beta$pic members, 30 Myr for Tucana-Horlogium members and 70 Myr for
AB Doradus members.

\subsubsection{2MASSJ044337.6+000205.2}
    This object is not known as young moving group member but is
    identified as a low gravity M9 by \citet{Cruz.2007} % et al.,ApJ, 2007
    and as having lithium absorption \citep{Reiners.2009}.% which   \& Basri (ApJ, 2009)
     This
    implies the object is a brown dwarf below lithium burning mass or
    a slightly higher mass star/brown dwarfs that has not yet burned
    all its lithium. Using BT-SETTL models, it appears that this
    combination of spectral type and lithium absorption is only
    possible for an object younger than $\sim$120Myr. We therefore
    present detection probabilities both for a young age hypothesis of
    50 Myr and an old age hypothesis of 120Myr.\\

\subsubsection{HIP114046/Gl887}
    This object is not either known as young moving group member which makes
    its age determination more challenging. Indeed, even if this very
    nearby M2 star (3.3pc) is identified in 
    Simbad as a pre-main sequence 
    star, different age estimation methods give wildly different
    values.
	\citet{Pasinetti-Fracassini.2001} % et al. A\&A,2001
  CADARS catalog that
        compiles angular diameters gives very high radius estimates (respectively
    0.62 and 0.56R$_{\odot}$), derived by \citet {Lacy.1977,Johnson.1983}
    %&Wright, 1983
 from parallax and visual
    apparent magnitude. According to BT-SETTL models this would mean
    HIP114046 is aged of about 30-50 Myr. However the corresponding
    near-infrared colours would not fit the 2MASS colours, and the near-infrared
     magnitudes would be more than 1 magnitude too bright. 
     Using $V,J,H,K$ magnitudes and the corresponding colours together
     with its spectral type of M2
     ($\sim$3700K)  and comparing with BT-SETTL models, we derive that
     HIP114046 is probably a 
     0.5-0.55M$_{\odot}$ star aged anywhere between
     100 Myr and 10 Gyr, with a radius of approximately
     0.35R$_{\odot}$. The strong discrepancy in radius with earlier
     work is difficult to explain but could be linked to the strong improvement of M dwarfs
     models in the last 30 years. According to \citet{Ehrenreich.2011}, this 
     object is not very active in X ray ($R_X=-5.13$), giving more weight to the
     hypothesis that the object is not very young. The high absolute
     proper motion of this star, also points toward an older age. On
     the opposite, the detection of lithium absorption
     \citep{Torres.2006} favours a
      very young age (below $\sim$30Myr), that is difficult to
      reconciliate with its apparent luminosity.

     We try in the following to bring more quantitative constraints on the
     age of HIP114046 using gyrochronology \citep[see][for
       instance]{Barnes.2003}. There are $vsin(i)$ measurements of
     rotational velocities on 
     the star of 1.0$~km.s^{-1}$ by \citet{Nordstrom.2004},    %et al., A\&A,
     where $v$ is the rotational velocity at the equator and $i$ is
     the inclination of the
     axis of rotation with respect to the line of sight. Assuming the
     diameter of 0.35R$_{\odot}$ given by BT-SETTL models for
     HIP114046, and assuming various values for $i$, we used
     the relations between rotational periods and age given by
     \citet{Delorme.2011}  to derive the ages listed in table
     \ref{HIPage}. This indicates that HIP114046 is probably younger
     than 1Gyr and is also compatible  with a much
     younger age. We cannot derive a lower limit on the age of
     HIP114046 because the $i> 45 \dg$ hypothesis brings the rotation
     period in the range of relatively fast M dwarfs rotators that
     have not converged yet toward the clean age-period relation on
     which gyrochronology is grounded. Within
     this hypothesis, HIP114046 would be younger than the 
     convergence time for early M dwarfs, but could not have its age
     more accurately measured by gyrochronology.  Since the
     convergence time for early M-dwarfs has been measured  by
      \citet{Delorme.2011,Agueros.2011} to % and  Ag{\"u}eros et al., ApJ, 2011
     be about the age of the Hyades \citep[$\sim$625
       Myr][]{Perryman.1998}, $i> 45 \dg$ would mean HIP114046 could 
     have any age 
     between 0 and $\sim$625 Myr. In this article we 
    present detection probabilities of HIP114046 both for a young age
    hypothesis of 
    100 Myr and a conservatively old age hypothesis of 2000Myr.

\begin{table}
 \caption{Rotation periods and corresponding Gyrochronological ages derived for
   HIP114046 assuming different value for the inclination angle
   $i$. Errors are derived using a 10\% error on the $v.sin~i$
   measurement and the dispersion around the age/period relation
   observed in the Hyades by \citet{Delorme.2011}. \label{HIPage}}
\begin{tabular}{c|c|c} \hline \hline
$i$(deg.)& Rotation Period(days) & Age(Myr) \\ \hline
0& 17.7 & 1060$\pm$200\\ \hline
30& 15.3 & 820$\pm$150 \\ \hline
45& 12.6 & $<$625\\ \hline
60& 8.9 & $<$625\\ \hline
\end{tabular}
\end{table}

\section{Data analysis} \label{data_analysis}
\subsection{Image reduction}
All images were reduced using the IDL pipeline developed for AO reduction at the Institut
de Plan\'etologie et d'Astrophysique de Grenoble (IPAG), that we
describe in the following. 

\subsubsection{Bad pixels removal, flat-fielding and recentering}
To achieve optimal flat-fielding and bad pixel removal we acquire sky
flats for each detector/dichroic/filter combination used for each
run. The raw data are flat-fielded and bad pixels are removed
using these twilight flat images.
The background for each data cube is calculated from the median of
  all cubes observed within 250s of the reduced cube and subtracted
  to each frame within a cube. The quality of these individual reduced
  frames is  
  assessed to remove frames for which the AO loop was open
  or the AO correction was poorer than usual. This automatic
  selection process makes use of the cubes statistics of flux peak and
  total flux in the image to remove these lower quality frames.  
The remaining good-quality intra-cube frames are then
accurately recentered using Moffat fitting, before each cube is
collapsed to produce one image. Each of these collapsed image is then
recentered again using Moffat fitting to ensure they all are centred
at the same position.

\subsubsection{Additional reduction steps for Angular Differential Imaging}
  Seven of our targets were observed with the rotator off and at an
  hour angle that allowed sufficient field rotation (typically $\geq
  25 \dg$) to use simple ADI reduction method \citep{Marois.2006} to
  remove the PSF of the central star.

  We used several advanced variations on the ADI
  technic on each of these targets, namely smart ADI \citep{Lagrange.2010}, LOCI \citep{LafreniereLOCI.2007}, and 
  the new slightly different method that we describe here, ``weighted"
  ADI. Weighted-ADI is a variant of ``smart" ADI. The references used
  to build the smart-ADI PSF for a given image are chosen with the constraint
  that the field must have rotated by a given angle with respect to
  the image, to mitigate the self-subtraction of possible
  companions. The $n$ references closest in time to the image, with $n$ being a free
  parameter that we usually set to 10, are then median-combined to
  produce the PSF  that will be subtracted to the image.  
Since the PSF evolve with time, a natural step beyond smart ADI is to
weight each reference by a value related to the invert 
of the time-span between image and reference before combining
them. We used a weighted-ADI method  working exactly as a smart
  ADI but for the use of $\frac{1}{\Delta t}$ as a
weight for each reference.
 As can be seen in table \ref{detlim_ADI} the method yields results that are
 slightly better compared to smart-ADI, and particularly interesting
 in case of moderate on-sky rotation ($\leq$30$\dg$).

\subsubsection{Additional reduction steps for classical imaging}
   Ten of our targets were observed in classical imaging mode, with
   field tracking on. For these we also used independently several subtraction methods,
   namely low frequency spatial filtering (a sliding median filter
   with a box size of 4$\times$FWHM), subtraction of radial profile
   (for a given annulus centred on the primary, we subtract the median value of all pixels
   within the annulus) and subtraction of the image rotated by $\pi$.
% or with field tracking of, but without enough
%   field rotation to carry out ADI reduction 

\smallskip
For all targets, whether or not observed in ADI, a stacked image of the full NACO field of view,
extended by the offset of the dither pattern was produced to search for
background-limited  large separation
companions. Since most of our observations were windowed to 512x512
pixels, the resulting field was 19.5$\times$19.5$\arcsec$, typically
400\,AU by 400\,AU on sky. In
the case of 2M1207, the full 1024$\times$1024 frame was read,
resulting in a fourfold increase in the research area.

 \subsection{Companion detection tools}
\subsubsection{Automatic check for candidates}
Given the different possible signatures of a companion in residual
images of AO observations once reduced with the various
ADI technics, the trained human eye is usually the most efficient
way of looking for companions in the speckle-dominated region close to
the star. See Fig. \ref{GJ3305} for an example
of a target with a confirmed and a possible companion after ADI
subtraction. However, searching by eye lacks objectivity and does 
not allow to obtain quantitative value of the threshold above which
potential companions are investigated. Additionally, in the
background dominated region far from the central star, automatic
routines can be more reliable than eyeballing to systematically look
for low Signal to Noise (hereafter SN) companions in this relatively large area.

 In order to have an objective way of looking for companions in our
final residual image we built an automatic detection routine that works as follows:
\begin{itemize}
\item We create a detection map which is the residual image
  median-boxed on 3 pixels. 
\item The successive maxima of this detection map are investigated as
  a potential companions. This detection image is not used
  afterwards but enables to distinguish between detections caused by
  remaining bad pixels or similar artefact from more reliable detections.
\item For each maximum the local background and noise are estimated
  as the median and the dispersion of pixels values 
within the intersection of 2 annuli in the initial residual image, one centred around the central star and another around the candidate itself.
\item The resulting SN of the candidate on
  1.5.$\times$FWHM-diameter aperture is derived, and is tested against a user-defined SN
  threshold.
\item  If this test is passed, a Moffat fit on the
  core of the signal is carried out. This fit is severely constrained so
  that it does not differ wildly from the expected FWHM-wide Moffatian
  of the core of a true PSF and mitigate the effects of ADI
  self-subtraction, as described below.
 \item A new SN, using the flux of the Moffatian fit under a
  1.5$\times$FWHM-diameter aperture centred on the peak of the flux, is
  derived.
\item If this SN is higher than a given threshold, the candidate
  is pinpointed as a potential companion that needs to be confirmed
  by a visual inspection complemented by regular outputs from the fit,
  such as the $\chi ^2$.
\end{itemize}

   The rationale we used to mitigate the effects of
   ADI self-subtraction is the following. Firstly, the fit is carried out
   on a small aperture, typically within the first Airy ring. This
   excludes most of the negative ADI-signatures of the companion (see
   Fig. \ref{fp_image}) which would
   mathematically lower the signal of a companion while they
   actually enhance our confidence in a detection. Secondly, the fit
   is weighted to artificially give more weight to the central pixels
   which are less affected by self-subtraction.

\subsubsection{Visual check for candidates}
  Though the fit produces useful information regarding the shape of
  the detected sources, it is not reliable enough, especially in the
  speckle-dominated area, to use as a tool
  for final candidate selection. Our strategy was therefore to assist human
  inspection of residual images with the detection routine, which
  provided a map of the 3$\sigma$ 
  automatic detections and their corresponding fit parameters. The
  subsequent human inspection of these sources allowed to retain only
  the most credible candidates, usually corresponding to a signal to
  noise higher than $\sim$5$\sigma$.

\subsubsection{2D detection limits maps}
   We took a particular care to derive accurate and meaningful
   detection limit map around each of our targets.

 The first step was
   to produce a noise map of each of our final residual image, by
   measuring the pixel noise within sliding square boxes of 5$\times$5 pixels
   (typically $\sim$1.5$\times$FWHM). This
   provides the noise per pixel and could be used directly to derive
   the peak value of a signal that would be above a given SN
   threshold. However, since actual detections do not rely on a 
   single pixel being above a signal to noise threshold, we preferred
   to derive the 5$\sigma$ detection limit on a 1.5$\times$FWHM-diameter
   aperture, comparing the noise within the aperture to the flux of
   the unsaturated PSF of each target within the same aperture. 

\begin{center}
\begin{figure*}
\includegraphics[width=18cm]{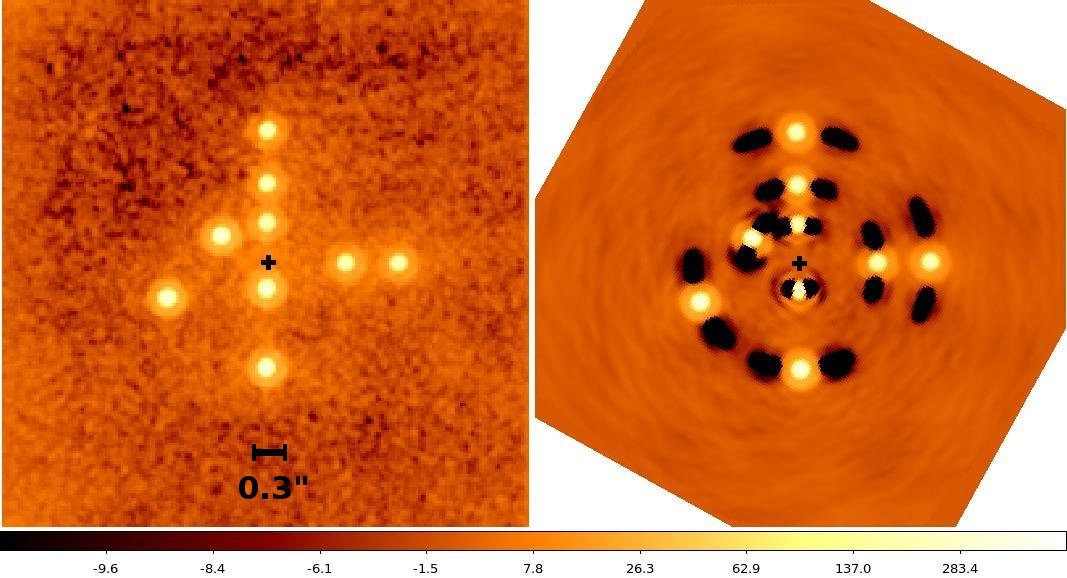}
\caption{{\bf Left:} False colour image of fake planets injected at
  separation ranging from 0.27$\arcsec$ to 1.35$\arcsec$in a cube
  with the same sky rotation as the science target. {\bf Right:}
  Residual image of these fake planets after weighted-ADI procedure
  were applied, simulating an on-sky rotation of
  29$\dg$. Self-subtraction effects are visible as negatives ``wings"
  around the fake planets. The rotation centre for the ADI procedure
  is marked by a black cross. A logarithmic scale is used.
\label{fp_image}}
\end{figure*}
\end{center}

In the case of ADI-reductions, a second step was to correct these detection maps from the flux loss
caused by self-subtraction. We created cubes
of images of high SN fake planets with the same field
rotation as the science images and applied the same ADI procedures as
used for the science images. We compared the flux injected to the
flux recovered by our detection pipeline on the final reduced fake
planets images (see Fig. \ref{fp_image}) to derive the actual flux
loss due to self-subtraction in ADI-processing on a range of radius
from the central star. Note that since these fake planet images are
used to measure only self-subtraction effects, the central star is not
included in the image. The effects of the central star residuals are
already taken into account in the noise map produced in the previous
step, where it dominates the local noise at short separations. 

\begin{figure}
\includegraphics[width=9.5cm]{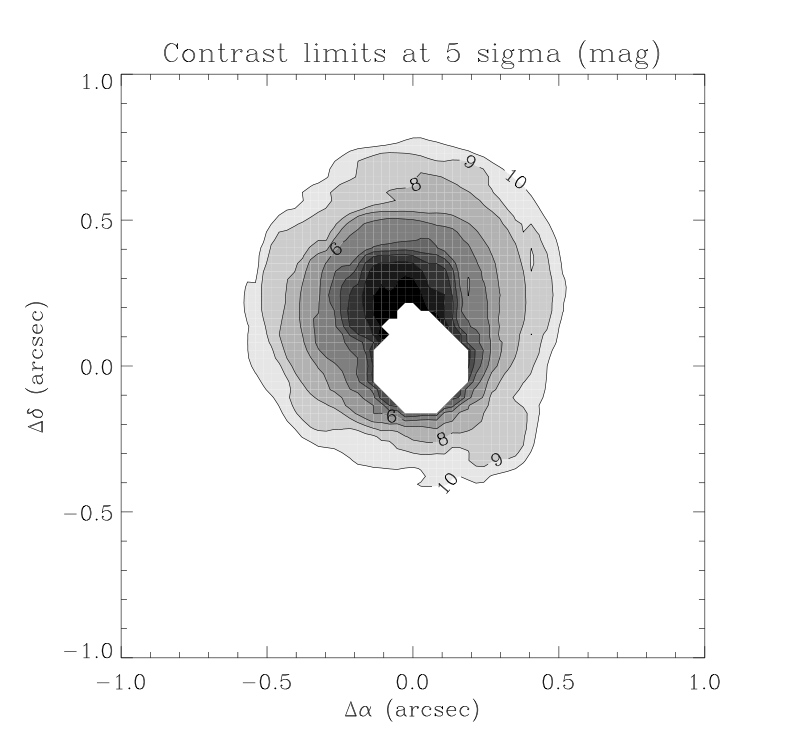}
\caption{ Final contrast map for the GJ3305AB system.
  Contours identify magnitudes of contrast at 5$\sigma$ achieved at a given
  position. The white area in the centre 
  shows the position of the primary where no contrast was measurable
  and dark zone slightly on the north is the higher noise
  area induced by GJ3305B. North is up and East is left.
\label{detlim_map}}
\end{figure}

 Finally, the detection limit in Arbitrary Detector Unit was converted
 into a contrast in magnitudes by scaling it to the measured flux of
 the unsaturated science target. Figure \ref{detlim_map} shows one
 instance of such a contrast map where the target is a close binary
 to illustrate the advantages of a 2D mapping approach over a 1D
 detection limit curve in a strongly asymmetrical case.

Such 2D contrast maps were produced for all targets and for all the
   reduction technics used. Tables \ref{detlim_ADI} and
   \ref{detlim_nonADI} give the
   median of the contrast achieved in each image both for a close
   separation area ($<$18.5 pixels=0.5$\arcsec$) and a wider
   separation area ($>$18.5 pixels and $<$100
   pixels=2.7$\arcsec$). Inner pixels for which there was not enough
   rotation to find a suitable PSF reference either in LOCI or sADI/wADI were
   removed from the statistics of all reduction modes. Computing values over area rather thin
   annuli, and using the same pixels for each set of images, provides
   robust statistics to   
   quantitatively compare different reduction technics. % (see table
                                % \ref{ADIparameters}). 
    Note that these tables present detection limits obtained with a
   single set of ADI and LOCI parameters and that slightly better detection
   limits (typically by $\leq$0.3mag ) can be obtained by manually fine tuning
   the parameters for each target, especially when using LOCI. We carried
   out such optimised analysis each time we had a hint that a potential
   candidate companion could be visible, but since the overall
   sensitivity gain was modest, we choose to present the more
   self-consistent detection limits obtained with a single set of
   parameters for all targets. 

\begin{table*}
 \centering
 \caption{Median contrast achieved for ADI-compatible observations,
   ranked by field-rotation, at
   5 $\sigma$ limit for short($\leq 0.5 \arcsec $) and large
   separation areas. Best values are highlighted in bold. \label{detlim_ADI}}
\begin{tabular}{c|c c|c c |c c|c c|c} \hline \hline
Target Name & \multicolumn{2}{c|}{Simple ADI} & \multicolumn{2}{c|}{Smart ADI} & \multicolumn{2}{c|}{Weighted ADI} & \multicolumn{2}{c|}{LOCI}& Sky Rotation\\
            & $\leq 0.5 \arcsec $  & $> 0.5 \arcsec $  &$\leq 0.5
\arcsec $  & $> 0.5 \arcsec $ &$\leq 0.5 \arcsec $  & $> 0.5 \arcsec
$   &$\leq 0.5 \arcsec $  & $> 0.5 \arcsec $  & (Degree) \\ \hline
BD01 2447       & 9.1$\pm$0.15 &  {\bf12.69$\pm$0.02} & 9.3$\pm$0.11  & 12.46$\pm$0.03 & {\bf 9.9$\pm$0.14}  & 12.53$\pm$0.03 &  9.7$\pm$0.11  & 12.09$\pm$0.02 &
23 \\ \hline
GSC0856        & 8.8$\pm$0.07 &  {\bf 9.90$\pm$0.02} & 8.7$\pm$0.08  & 9.70$\pm$0.02 & {\bf 8.9$\pm$0.08}  & 9.79$\pm$0.02 &  8.3$\pm$0.08  & 9.38$\pm$0.02  & 29\\ \hline
2MASS J0443     & 5.8$\pm$0.06  & {\bf 6.84$\pm$0.02} & 6.1$\pm$0.06  & 6.71$\pm$0.02 & {\bf 6.2$\pm$0.06}  & 6.74$\pm$0.02 &  6.1$\pm$0.06  & 6.54$\pm$0.02 & 29  \\ \hline
GJ3305 (12/2009)         & 9.1$\pm$0.26 & {\bf 11.37$\pm$0.02 }& 8.9$\pm$0.27  & 11.24$\pm$0.03 & 9.0$\pm$0.28  & 11.30$\pm$0.03 &  {\bf 9.2$\pm$0.20} & 11.07$\pm$0.03
& 40 \\ \hline
HIP114046       & 9.6$\pm$0.15  & {\bf 14.23$\pm$0.03} & 9.9$\pm$0.15  & 14.04$\pm$0.03 & 9.8$\pm$0.17  & 14.05$\pm$0.03 &  {\bf 10.1$\pm$0.10}  & 13.34$\pm$0.04 &47 \\ \hline
GJ3305 (09/2011)         & 9.6$\pm$0.27  & {\bf11.47$\pm$0.03 }& 9.1$\pm$0.20  & 11.15$\pm$0.03 & 9.3$\pm$0.20  & 11.23$\pm$0.03 &  {\bf 9.7$\pm$0.21}  & 11.21$\pm$0.04
& 50 \\ \hline
HIP102409/AUmic & 9.4$\pm$0.15  & {\bf 13.16$\pm$0.03} & 8.8$\pm$0.14  & 12.90$\pm$0.03 & 9.1$\pm$0.17  & 12.95$\pm$0.03 &  {\bf 10.2$\pm$0.10}  & 12.73$\pm$0.04 &
74 \\ \hline
TYC7443-1102-1  & {\bf 9.3$\pm$0.09}  &  {\bf 9.84$\pm$0.02} & 8.9$\pm$0.09  & 9.72$\pm$0.02 & 9.0$\pm$0.08  & 9.78$\pm$0.02 &  8.6$\pm$0.07  & 9.33$\pm$0.02 & 95 \\ \hline
\end{tabular}
\end{table*}

% special GJ3305: companion exclu de la zone de calcul et
% normalisation centree sur le compagnon.

\begin{table*}
 \centering
 \caption{Median contrast achieved at 5 $\sigma$ limit over a short($\leq
   0.5 \arcsec $) separation area and a large separation area for
   targets observed with 
   classical AO imaging (without ADI mode). \label{detlim_nonADI}}
\begin{tabular}{c|c c|c c} \hline \hline
Target Name & \multicolumn{2}{c|}{Circular Profile Subtraction} &
\multicolumn{2}{c}{Subtraction of $\pi$-Rotated  Image} \\
            & $\leq 0.5 \arcsec $   &  $> 0.5 \arcsec $  &$\leq 0.5
\arcsec $   &  $> 0.5 \arcsec $ \\ \hline
V1005 Ori       & 7.5$\pm$0.18  & 10.28$\pm$0.03 & 7.0$\pm$0.15  & 9.89$\pm$0.02 \\ \hline
CD-571054       & 7.5$\pm$0.12  & 11.52$\pm$0.03 & 7.1$\pm$0.10  & 11.25$\pm$0.03 \\ \hline
BD-21 1074A       & 7.6$\pm$0.14  & 11.11$\pm$0.03 & 7.2$\pm$0.14  & 10.83$\pm$0.03 \\ \hline
%HIP30034         & 7.53  & 9.80 & 7.10  & 9.42 \\ \hline
 2M1139       & 5.5$\pm$0.10  & 5.80$\pm$0.02 & 5.1$\pm$0.09  & 5.47$\pm$0.02 \\ \hline
2MASS1207       & 6.5$\pm$0.09  & 6.65$\pm$0.02 & 6.1$\pm$0.14  & 6.34$\pm$0.02 \\ \hline
WWPsA        & 7.6$\pm$0.14  & 10.12$\pm$0.02  & 7.1$\pm$0.12  & 9.74$\pm$0.02  \\ \hline
 TX PsA       & 7.3$\pm$0.12  & 9.13$\pm$0.02 & 6.7$\pm$0.11  & 8.75$\pm$0.02 \\ \hline
 TWA25      & 7.1$\pm$0.14  & 9.75$\pm$0.03 & 6.7$\pm$0.13  & 9.37$\pm$0.02 \\ \hline
BD-136424         & 7.7$\pm$0.15  & 10.45$\pm$0.03 & 7.2$\pm$0.14  & 10.09$\pm$0.03 \\ \hline
\end{tabular}
\end{table*}

\section{Results}
 \subsection{Specially interesting targets}
We detail case by case a few specifically interesting objects, notably 2
 targets around which we identified candidate companions, none of
 which were physically associated with the central star.

\subsubsection{2M1139/TWA26}
   Our automatic detection routine identified a source 13$\sigma$
   above local background, at a separation $\rho$=13.4$\arcsec$ and a
   position angle $\theta$=196$\dg$ from 2M1139, at RA=11:39:50.8 and
   Dec=-31:59:34.8 

%After
%   looking on ESO archive we established that 2M1139 was observed by
%   Ginsky et al. on during observing run 084.C-0626(A) with SOFI at
%   NTT. We downloaded and reduced the data

% heu, ca fait moins bete de passer cela sous silence et de dire:
A check of 2MASS images shows a H=15.5 object at the position
of our $L'$ detection. Since this object was also detected in USNO
optical images, where it appears much brighter than the central star
it is obviously too blue and too bright to be a lower mass companion
of 2M1139. 
This was further confirmed by its proper motion which is not
compatible with the proper motion of 2M1139, at more than 10$\sigma$.

\subsubsection{The GJ3305~AB system}
\begin{center}
\begin{figure}
\includegraphics[width=8.6cm]{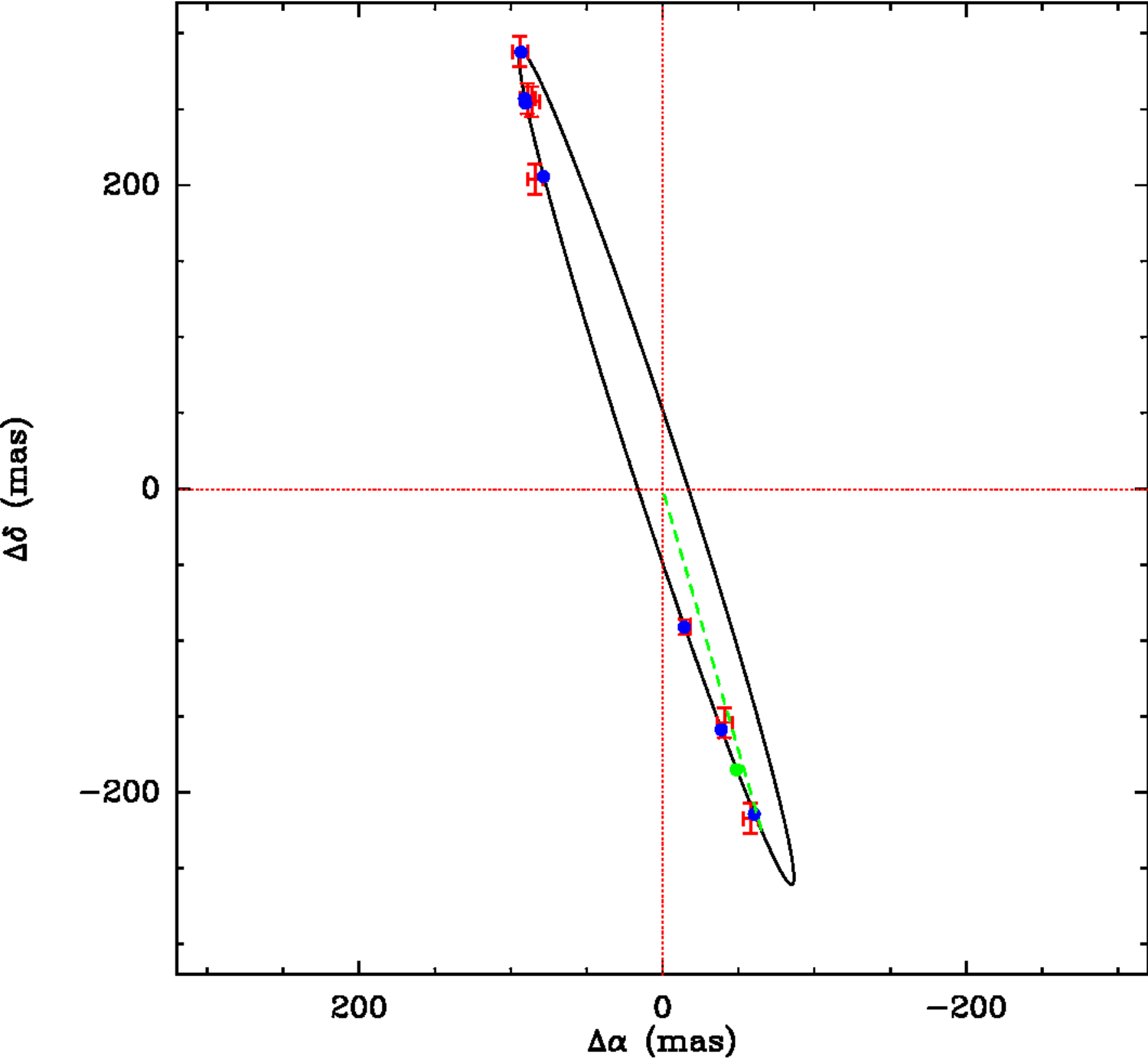}
\caption{ Best fit for GJ3305B orbit {\bf projected on the sky} in
  black, corresponding to a period of 21.5yr, and eccentricity of
  0.06, a semi-major axis of 8.5\,AU and an inclination of 93.2$\dg$. Observational
  astrometric points are in blue and the periastron is highlighted in green.
\label{orbit_fig}}
\end{figure}
\end{center}

\begin{figure*}
\includegraphics[width=18cm]{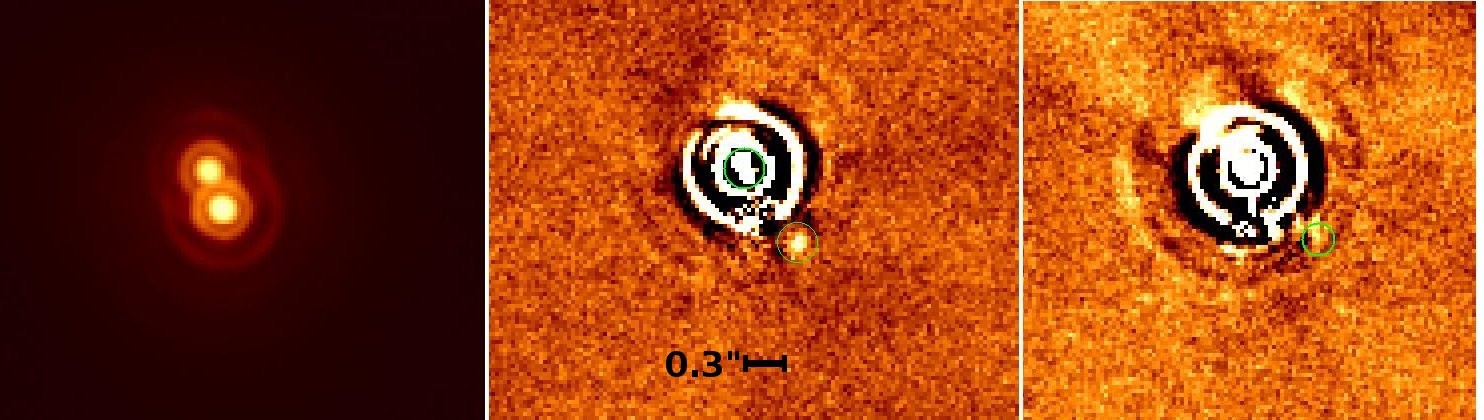}
\caption{{\bf Left:} Final stacked false-colour image of GJ3305AB from
  December 2009 data, in $L'$. North is up and
  East is left. Logarithmic scale is used. {\bf Center:} Residual image of GJ3305AB from
  December 2009 data, after GJ3305A is
  subtracted by simple ADI.  GJ3305B and the candidate companion are
  highlighted with green circles. Linear scale is used. {\bf Right:} Residual image of GJ3305AB from
  September 2011 data, after GJ3305A is
  subtracted by simple ADI. No source is detected at the 2009
  position of the 2009 source. The position expected for this source
  in case it was a background object is highlighted by a green circle. Linear scale is used. North is up and
  East is left.\label{GJ3305}}

\end{figure*}

\paragraph{GJ3305B.} 
 GJ3305 has been identified by \citet{Kasper.2007} as a low contrast
  close binary. Figure \ref{GJ3305} shows the reduced image side to
  side with the residual map after simple ADI subtraction. GJ3305B
  appears clearly in both images at a PA of 19.2$\dg$ and a separation of 0.27$\arcsec$. Its Airy rings are visible in the residual image up to the
  6th, a telltale sign of the effectiveness of the primary
  subtraction. By combining archive data of GJ3305AB shown on table
  \ref{orbit_tab} with our $L'$ observations we have 7 independent astrometric
  measurements of GJ3305B. Since they are relatively regularly
  spread on a significant fraction of the full orbit we attempted a
  Levenberg-Marquardt fit of an orbit for the system. We fixed the
  mass of this 12Myr-old $\beta$Pictoris moving group system to 0.85+0.5=1.35$M_{\sun}$
  using stellar evolution models from \citet{Baraffe.2003}. Though we
don't have enough astrometric points to derive accurate orbital
parameters, the fit consistently converged towards solutions where
the orbit is seen nearly edge-on (i$\sim$93$\dg$), with a low eccentricity
($e\sim 0.05$), with a resulting period of 20-25 years, and a
semi-major axis of 8-9\,AU, see Fig. \ref{orbit_fig}. Since only about one third of the
orbital period is covered by observations, the uncertainties are high
and it is not currently
possible to accurately constrain the mass of the system.

\begin{table}
 \caption{Position angle and separation of GJ3305B over the
   years. Typical separation error is 1mas and angle error is 1 deg. \label{orbit_tab}}
\begin{tabular}{c|c|c|c} \hline \hline
Date& Radius & Position Angle & Instrument-Filter\\
          & (mas) & ($\dg$) & \\ \hline
18/01/2003     & 225    & 195  & NACO-IB$_{2.09} ~ ~ ^1$ \\ \hline
08/01/2004  & 159  & 194      &  NACO-NB$_{2.12}$  \\ \hline
15/12/2004     & 93   & 189.5  & NACO-$L' ~ ~ ^1$ \\ \hline
17/11/2008    & 221  & 20.5  &  NTT/Astralux-$z' ~ ~ ^2$ \\ \hline
25/11/2009     & 269   & 18.6  & NACO-$L'$ \\ \hline
26/12/2009     & 272  & 19.2  &  NACO-$L'$\\ \hline
%mesure a 19.2 et correction sur l'orientation du detecteur de 
01/09/2011     &   303  &  18.1 &  NACO-$L'$ \\ \hline
\end{tabular}
\tablebib{$^1$ From \citet{Kasper.2007}.
$^2$ From \citet{Bergfors.2010}}
\end{table}

\paragraph{Candidate companion to GJ3305AB.}
 Another interesting source was identified around GJ3305AB in Dec. 26, 2009 images, at a PA of
 224$\pm 2 \dg$ and a radius of 0.38$\arcsec$ from GJ3305A. Since this
 source sits opposite to GJ3305B and between its Airy rings it is
 located in a relatively clean area of the residual map and the
 automatic detection routine identifies it as a 6 to 11 $\sigma$
 detection in all the different residuals
 maps produced with the different ADI technics.  It 
 was however  located very 
 close to the second Airy ring of the primary, and could be a
 speckle. Though it is likely this 2009 source was affected by speckle noise
 on the Airy ring, it was significantly above this local noise and the
 presence of ADI negative signatures around it supported it as
 real on-the-sky source rather than a speckle artefact.

 The resulting contrast
 with the primary in $L'$ is $\sim$9 magnitudes, far too faint to be
 detected at similar separations in all but one of the earlier dataset
 of GJ3305, the December 15th 2004, NACO $L'$ data 
 from which GJ3305B was initially identified. We retrieved the data from ESO archive and reduced it. Unfortunately
 given the lower exposure time and the now superseded ADI technics used
 for these observations, the noise at the position -assuming it is
 physically associated- of the faint
 candidate companion was higher than its expected signal and could not
 be used to confirm the detection. In the hypothesis this object would
 be a background source, the  proper motion of GJ3305
 is too low and not radial enough (RA=+46\,mas$.yr^{-1}$ and
 Dec=-65\,mas$.yr^{-1}$) to bring the 
 source out of the speckle dominated area, where it would have been
 detectable.  Since we could not either rule out this hypothesis on
 the basis of archive data, further observations were necessary.

 These observations were  carried out with NACO at VLT on September
 1st, 2011 and reduced following the ADI procedures described in
 section 3. The exposure time, rotation and the resulting detection
 limits for these new observations were slightly higher than those
 achieved on the 2009 data on which the candidate companion was
 identified. No source is detected in 2011  in the -small- area 
 compatible with the orbital motion of a 11\,AU planetary companion to
 GJ3305AB. This source is therefore not a planetary companion. A faint source
 is located close to the position expected if the 2009 detection was a
 background object. However the flux measured on this source in 2011
 is $\sim$0.7 magnitude fainter than the flux of the 2009 candidate,
 raising doubt on the physical nature of both detections. As a
 consequence, these data
 cannot establish whether the 2009 candidate was a
 background object or a speckle noise bump, but they allow to exclude
 it as a planetary companion.
 The residual images at both dates are shown on Fig.
 \ref{GJ3305}.

\subsubsection{AU Mic/HIP102409}
\begin{figure}
\includegraphics[width=9.4cm]{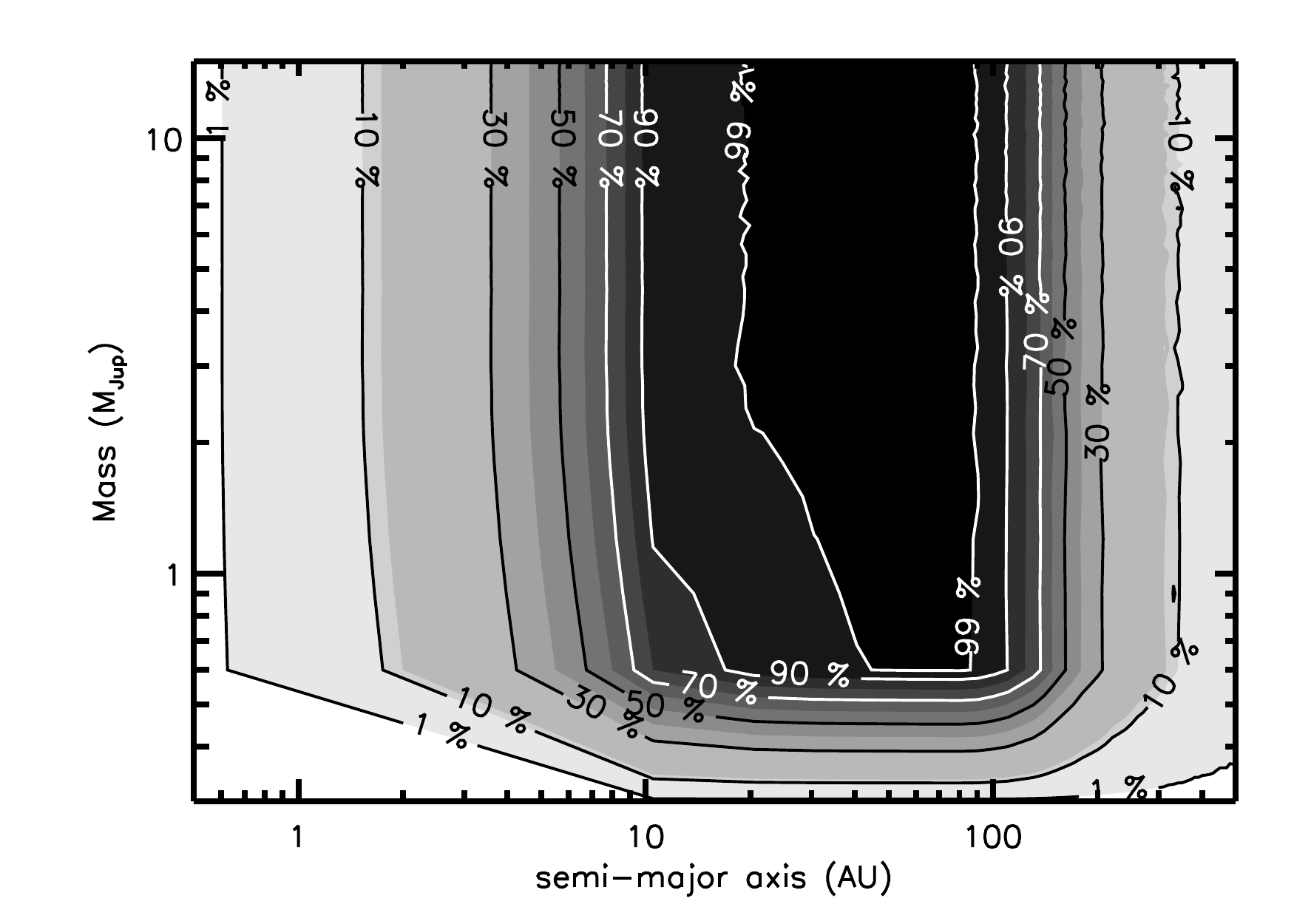}
\caption{ Planetary detection probability around AU Mic. Numbers on
  the contours indicate the detection 
  probability, taking into account the projection effects that could
  hid a planet in the line of sight 
   of the central star on a fraction of its orbit. The semi-major axis values are therefore in real AU and not in projected AU.
\label{detlim_AUMic}}
\end{figure}

  AU Mic. is known to harbour a edge-on debris disc \citep{Song.2002} which has
  been imaged using HST/ACS coronography down to 7.5\,AU
  \citep{Krist.2005}. Their HST images shows no significant trace of
  gravitational 
  perturbation of the disc by a large planetary body at close
  separation. We derived accurate detection probabilities as described
  in section 4.2 and show  that our data
  are able to exclude the presence of a planetary companion as light as
  0.6\mj beyond 20\,AU with $>$90\% confidence (see Fig.
  \ref{detlim_AUMic}), and improves the previous
 non-detections by \citet{Krist.2005} and \citet{Kasper.2007}.  Note
 that since our 
  detection limits show some sensitivity down to Saturn mass objects,
  below the lowest masses 
  available in the AMES-DUSTY model grid (0.5\mj), we extrapolated
  values between 0.3\mj and 0.5\mj, resulting in higher uncertainties
  in this mass range.

\subsection{Constraints on the existence of planetary mass companions}
 In order to evaluate the giant-planet detection probability around the
targets in our survey, we used the MESS code \citep{Bonavita.2011}.
For each star, the code generates 103 orbits per grid point in a mass
versus semi-major axis (SMA) grid with a sampling of 0.5 AU in SMA
axis and 0.5 Mjup in mass. 
The orbits are randomly oriented in space, and the eccentricities were
also randomly generated with a uniform distribution (with $e <$ 0.6).
For each random event, the code then evaluates
the projected separation and the position of the planet on the 
projected orbit at the time of the observation.  
With this approach we can take into account possible projection effects,
and then constraint the actual semi-major axis of the planets, instead
of the projected separation, while also using the whole spatial
information coming from the 2D detection limit maps evaluated in Sec.
3.2.3. The mass limits are obtained by translating these flux detection
limits into mass using the COND-base
models \citep{Allard.2001,Baraffe.2003} for temperatures below
1700 K and DUSTY-based models \citep{Chabrier.2000} for higher
temperatures.
The mass of the artificial planet is then compared to the detectable
mass at that position on the detection map, and the fraction of orbits
at each grid point that turn out to be detectable then corresponds to
the probability of detection at that grid point.

% In order to evaluate detectability taking into account the constraints
% of the possible orbits around each target star, we used
% the MESS code \citep{Bonavita.2011} to generate a grid
% of 50 by 50 values of masses and Semi-Major Axis (SMA). We adopted uniform
% distributions of mass, SMA and of eccentricities (with $e
% \leq 0.6$) rather than injecting distributions derived from RV, in a
% different separation range. The steps are 0.3 \mj in mass and 0.5\,AU
% in semi-major axis.
%For each point in the mass/SMA grid, the code produces 1000 possible orbits by
%randomly generating the other orbital elements. This allows to properly
%place each planets on the detector, according with its position on the
%projected orbit at the exact time of the observations. Then the planets
%mass is compared with the corresponding limiting value obtained from the
%images, and the fraction of orbits that are detectable at each given
%grid point is calculated. 

%In the peculiar case of the binary system GJ3305AB, we also excluded planetary orbits which
%would be instable given the orbital constraints described in Section
%4.1.2.

\begin{figure*}
 \caption{Detection probability as a function of the mass and semi-major axis of
the planets obtained with the MESS code (see text),  for all the
targets. \label{detlim_MESS_map}}
\begin{tabular}{c|c|c} 
\includegraphics[width=5.5cm]{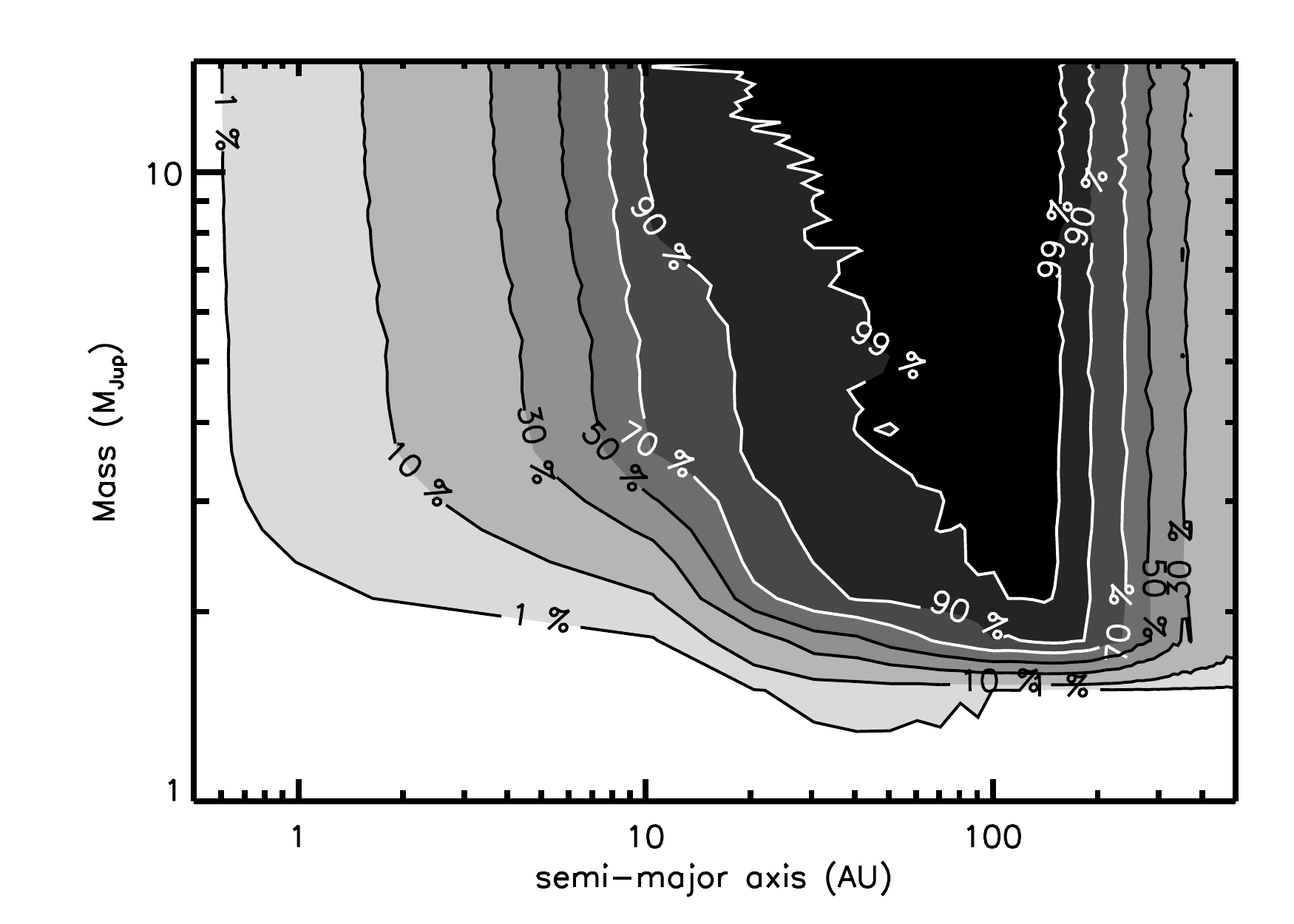}  &
\includegraphics[width=5.5cm]{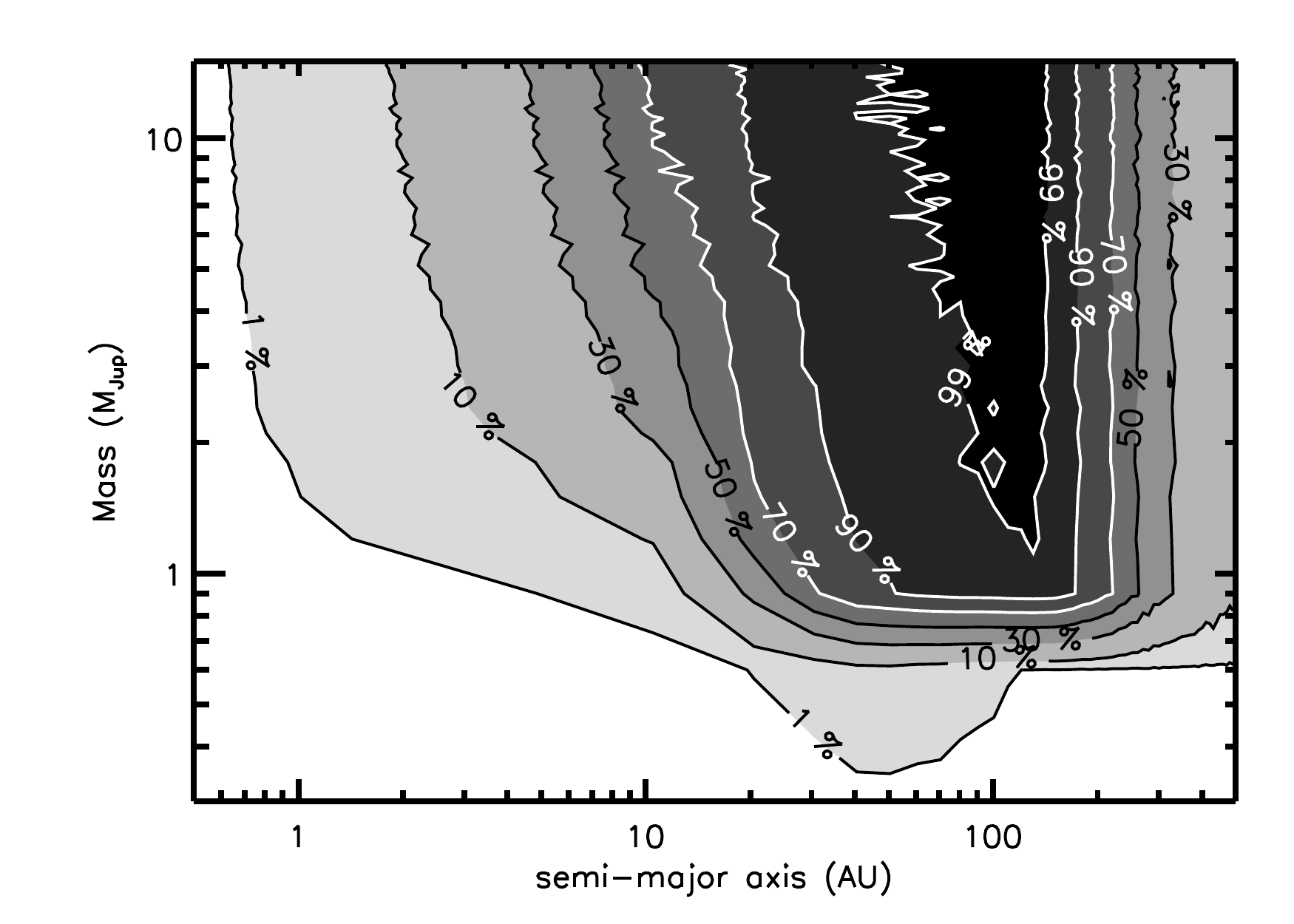} &\includegraphics[width=5.5cm]{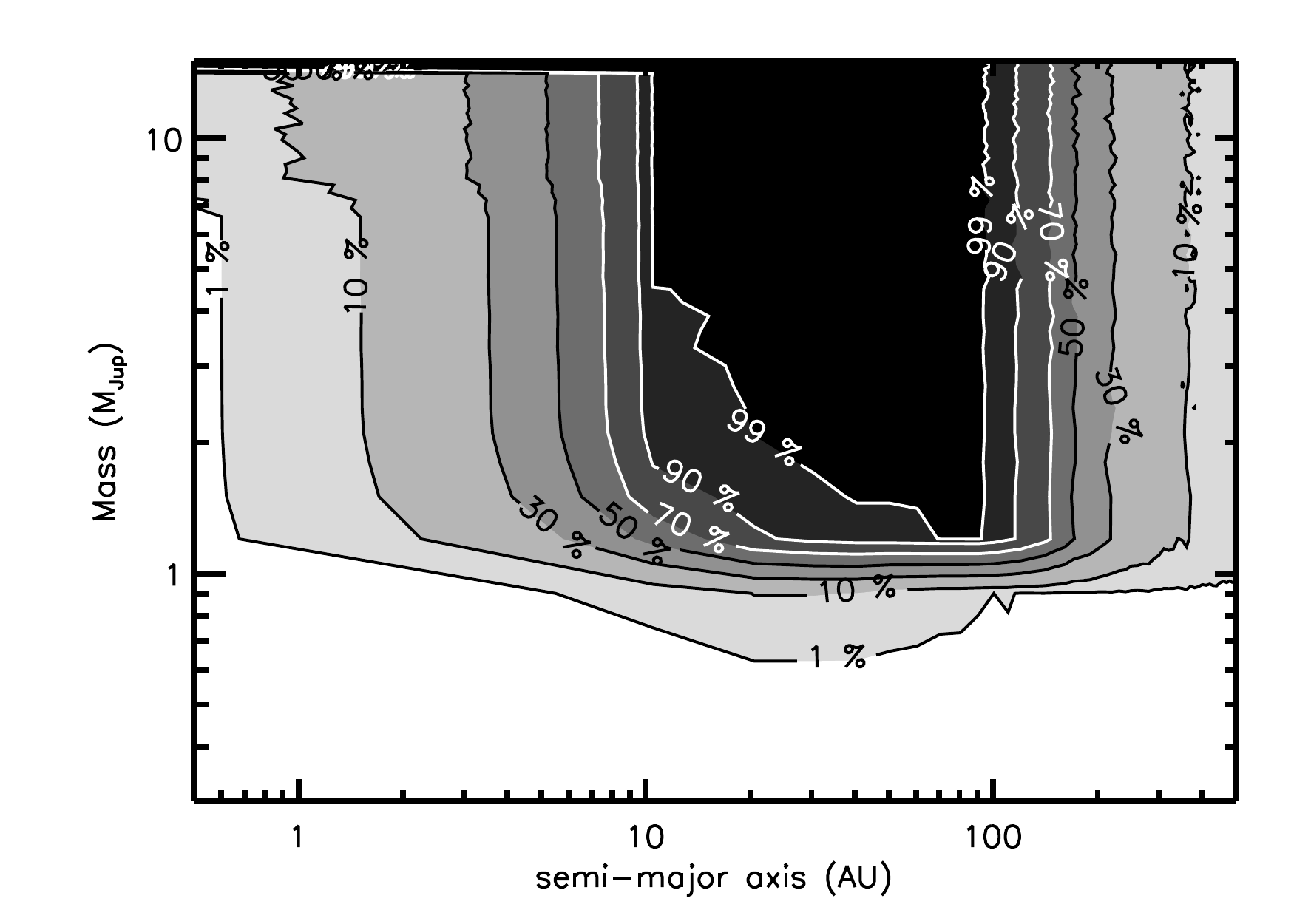}  \\ 
GSC0856 &GJ3305 (09/2011)& 2M0443 (if aged 50Myr) \\ \hline
\includegraphics[width=5.5cm]{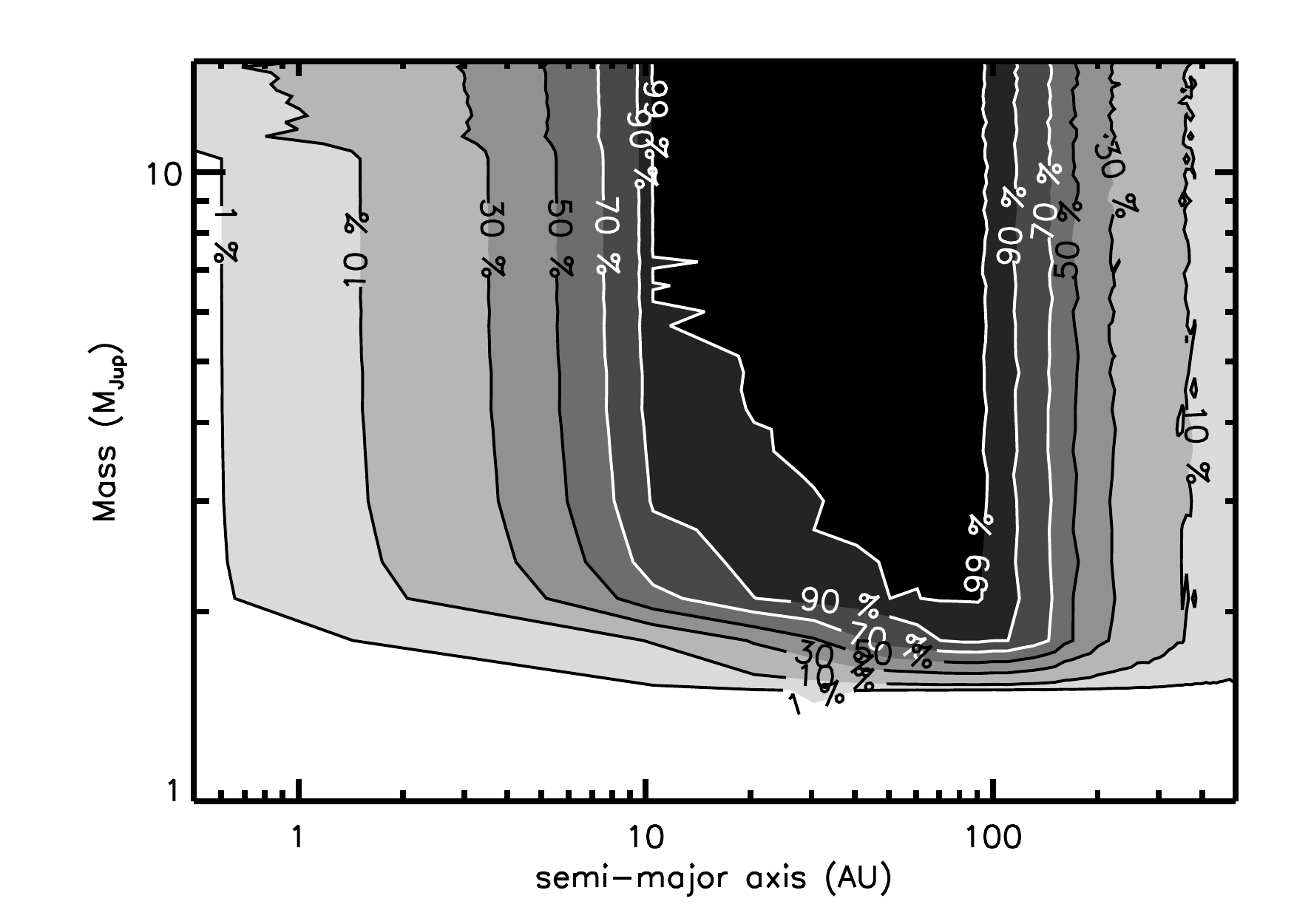}  &
\includegraphics[width=5.5cm]{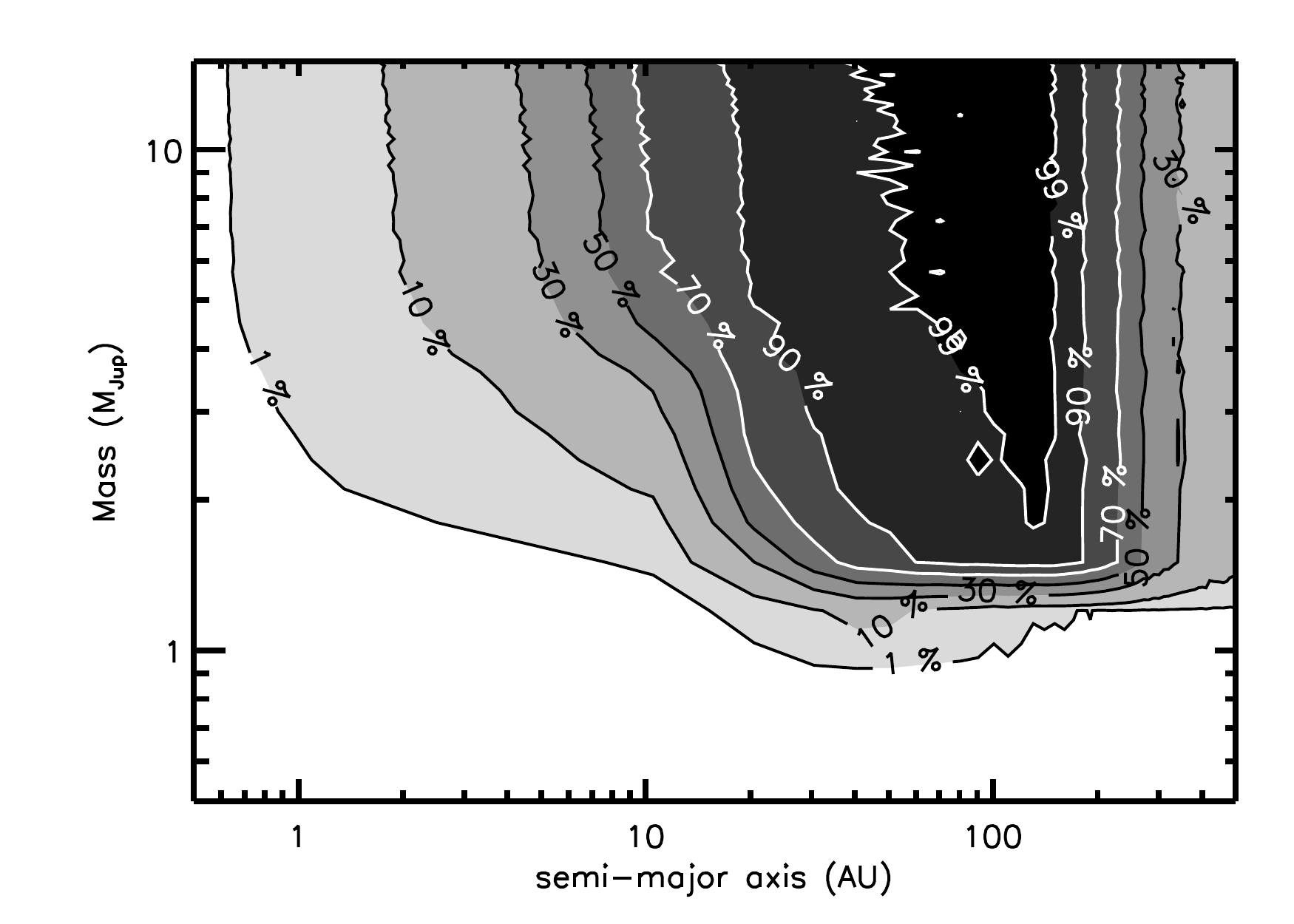} &\includegraphics[width=5.5cm]{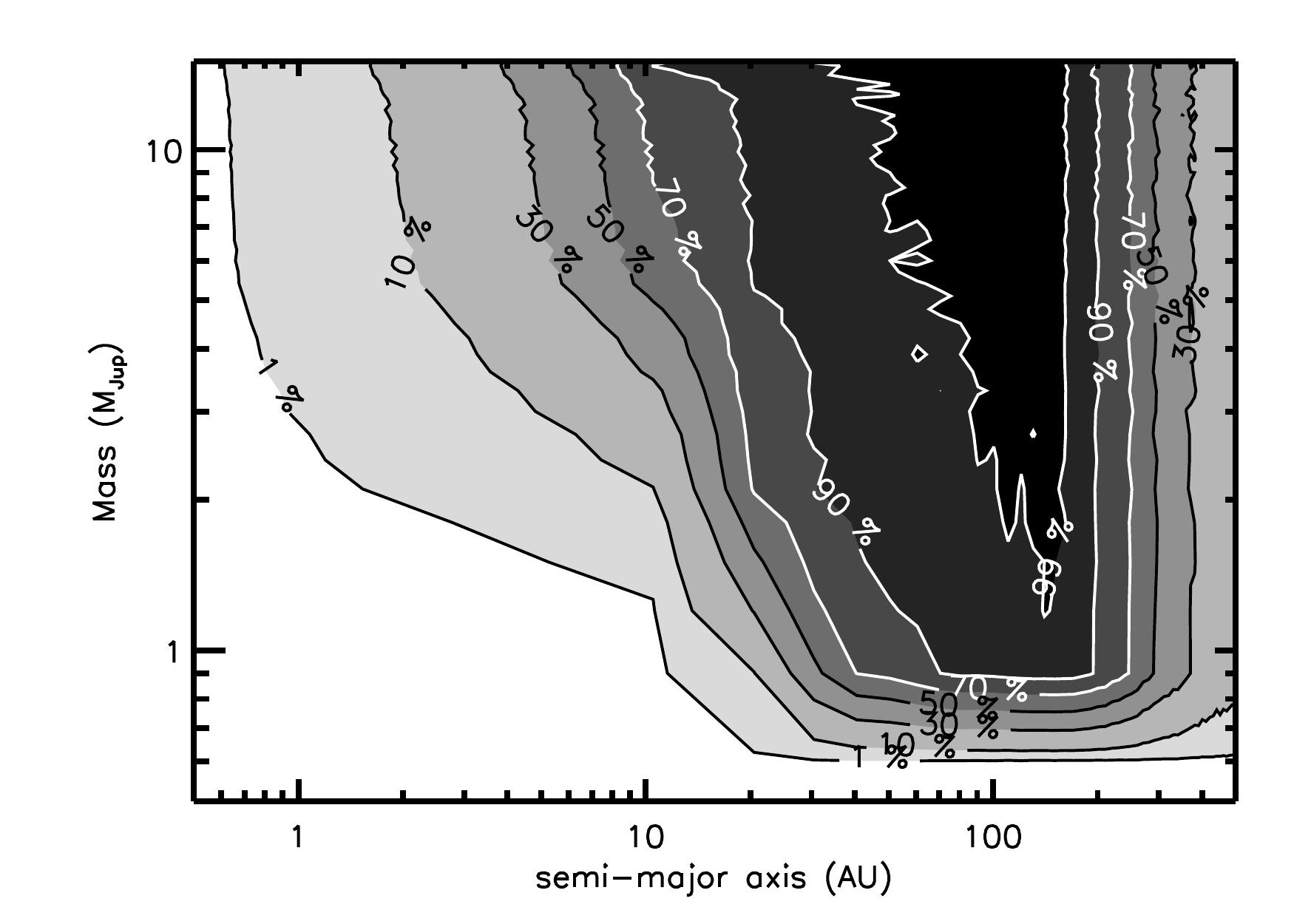}  \\ 
2M0443 (if aged 120Myr) &V1005 Ori& CD571054\\ \hline
\includegraphics[width=5.5cm]{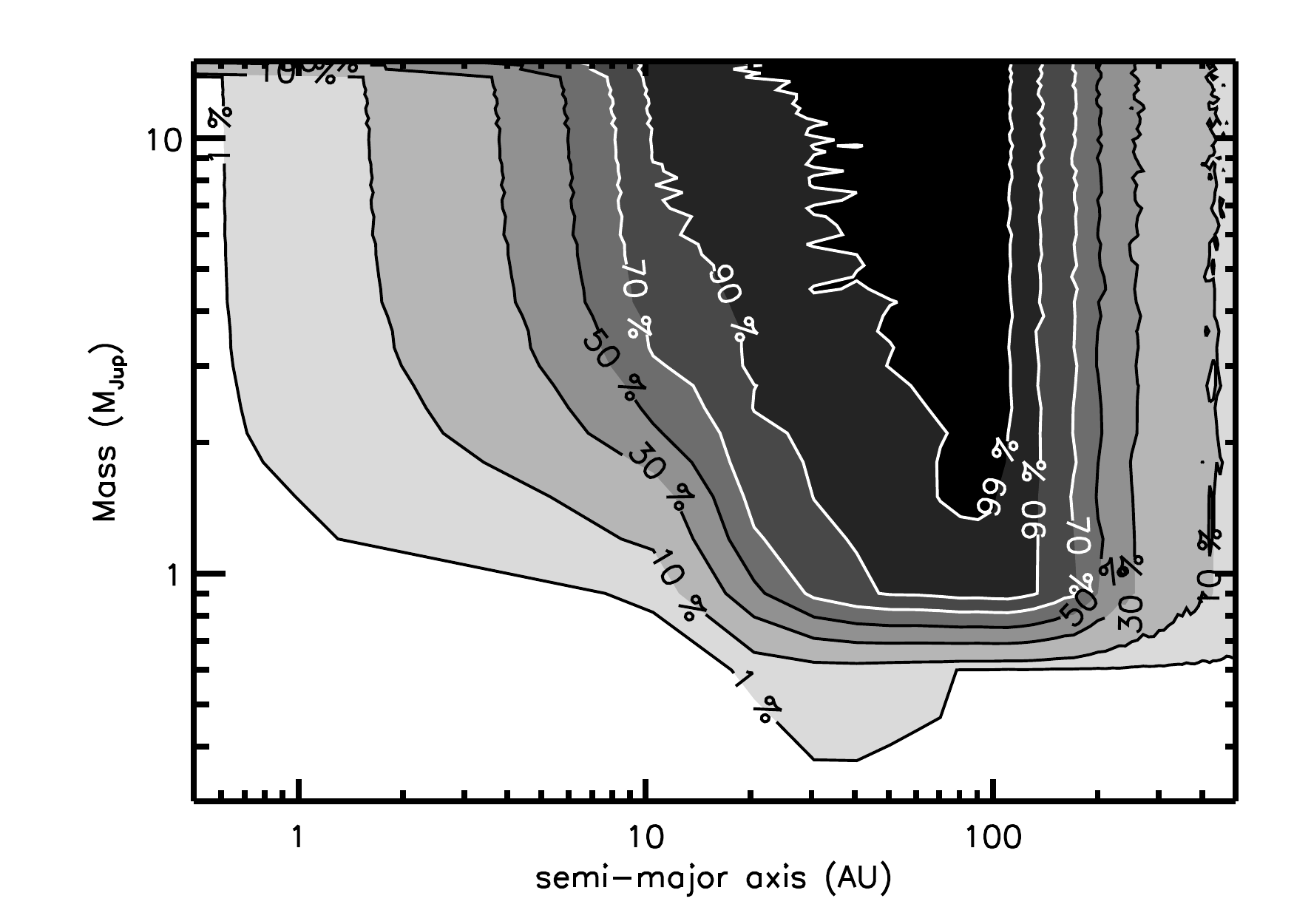}
  & \includegraphics[width=5.5cm]{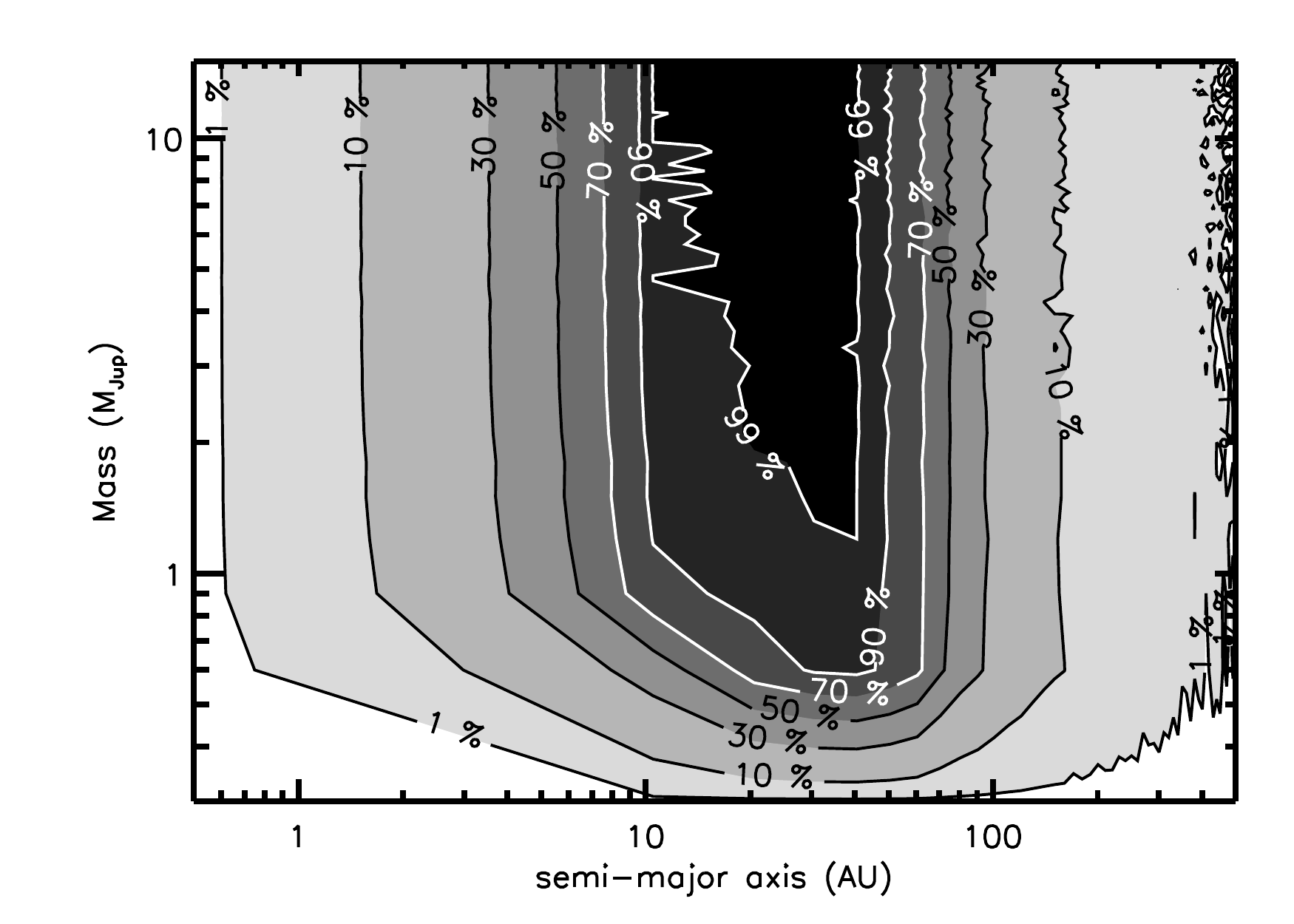} &\includegraphics[width=5.5cm]{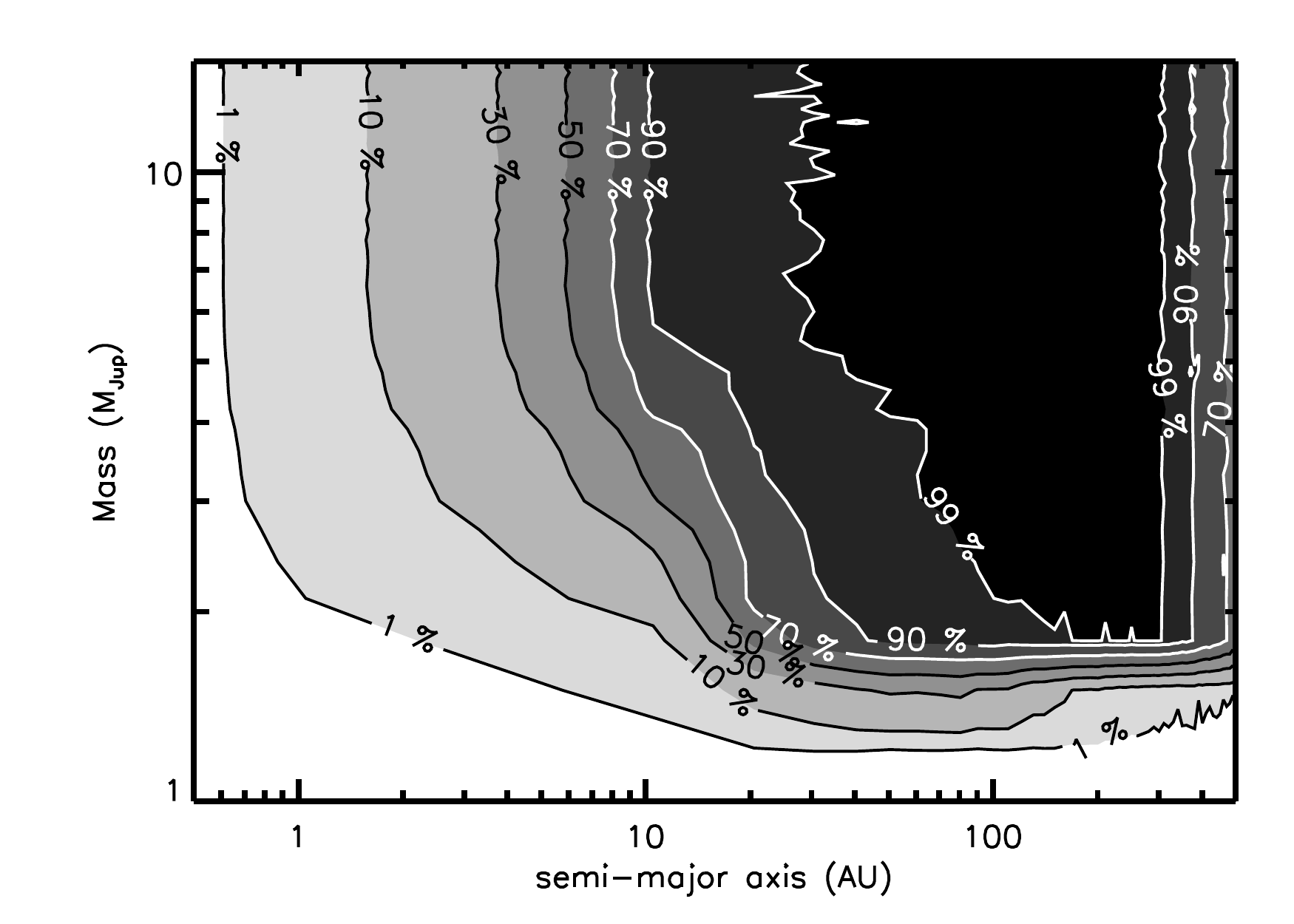}  \\ 
BD-21 1074A &BD01 2447& 2M1139 \\ \hline
\includegraphics[width=5.5cm]{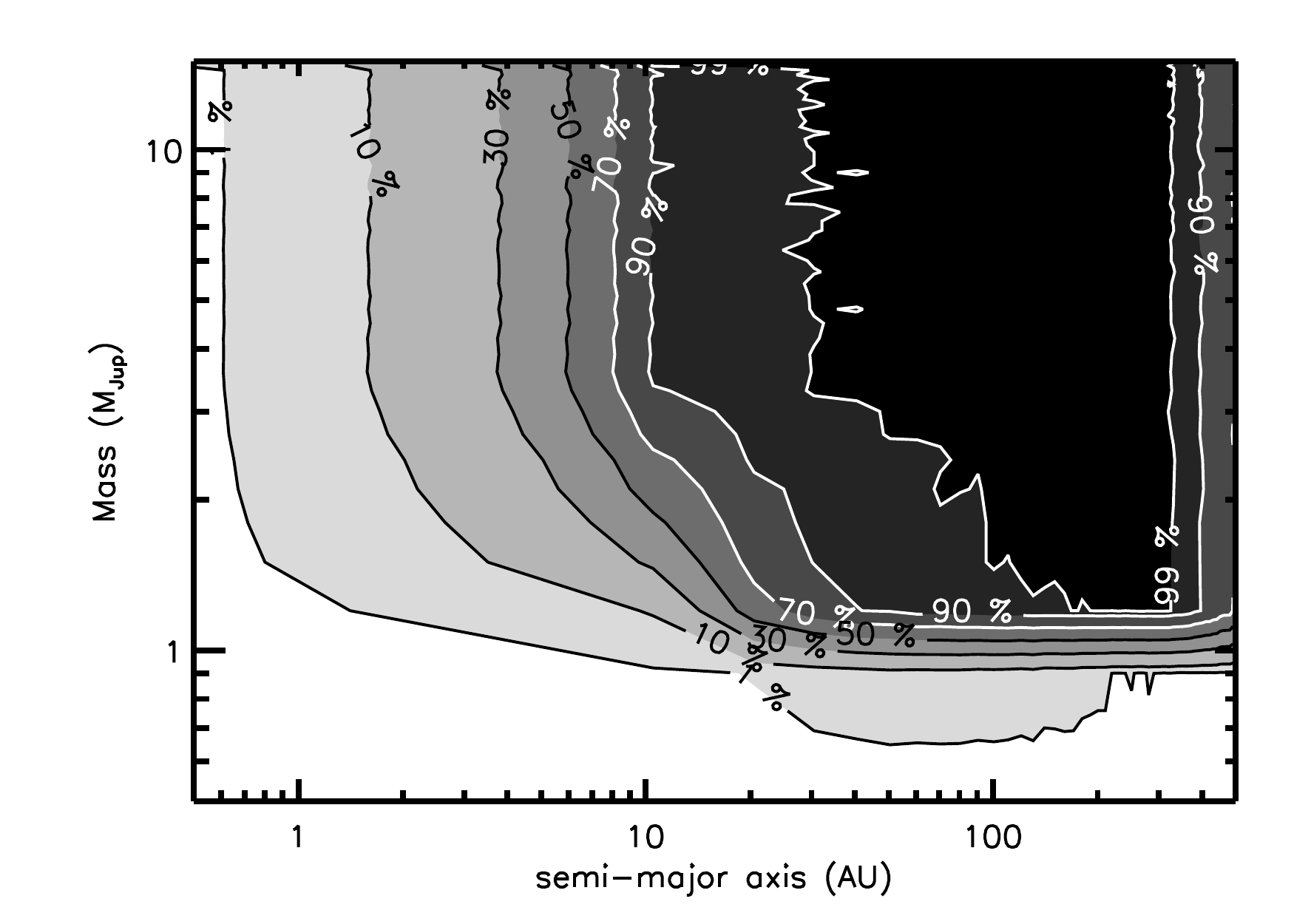}  &
\includegraphics[width=5.5cm]{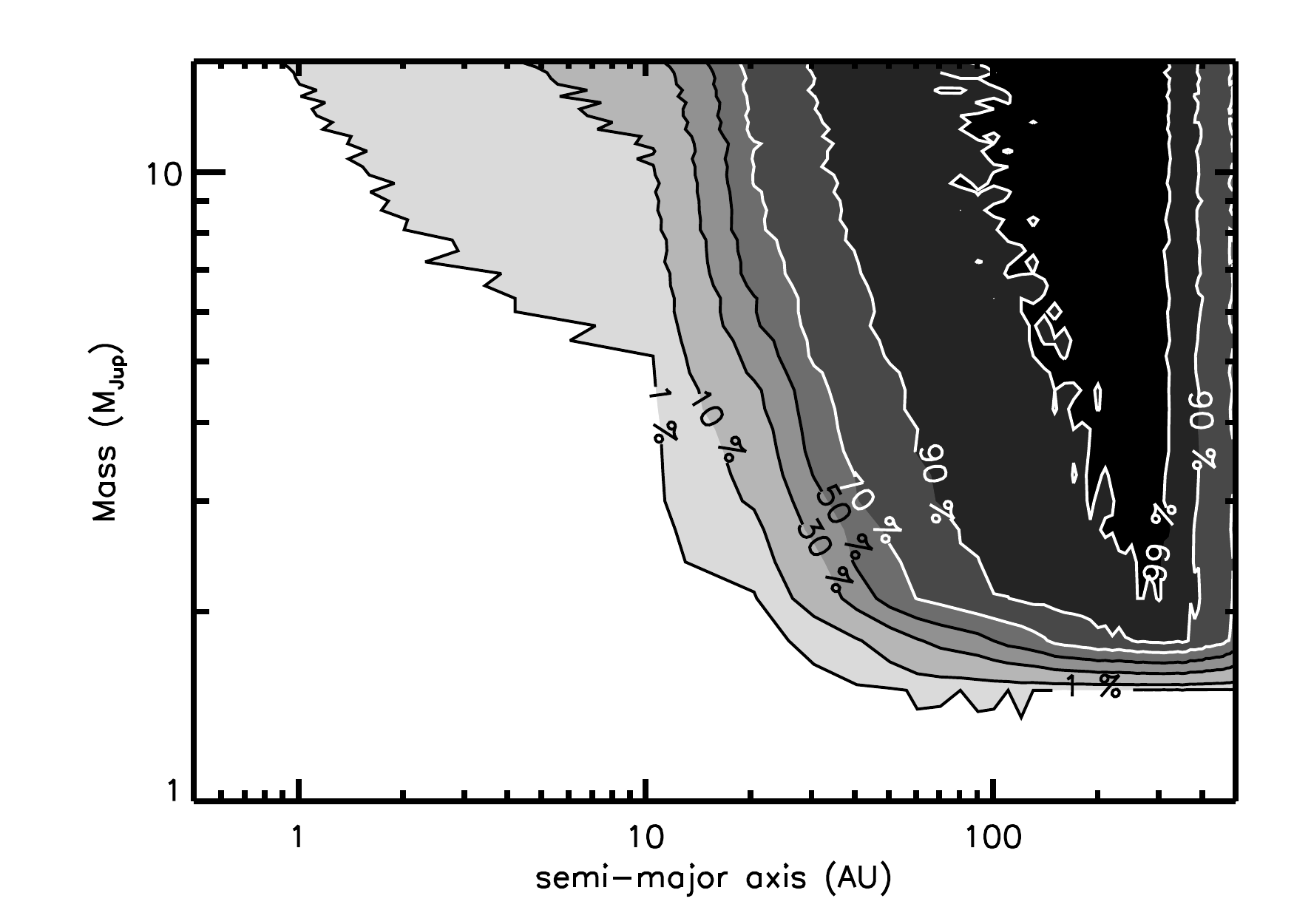} &\includegraphics[width=5.cm]{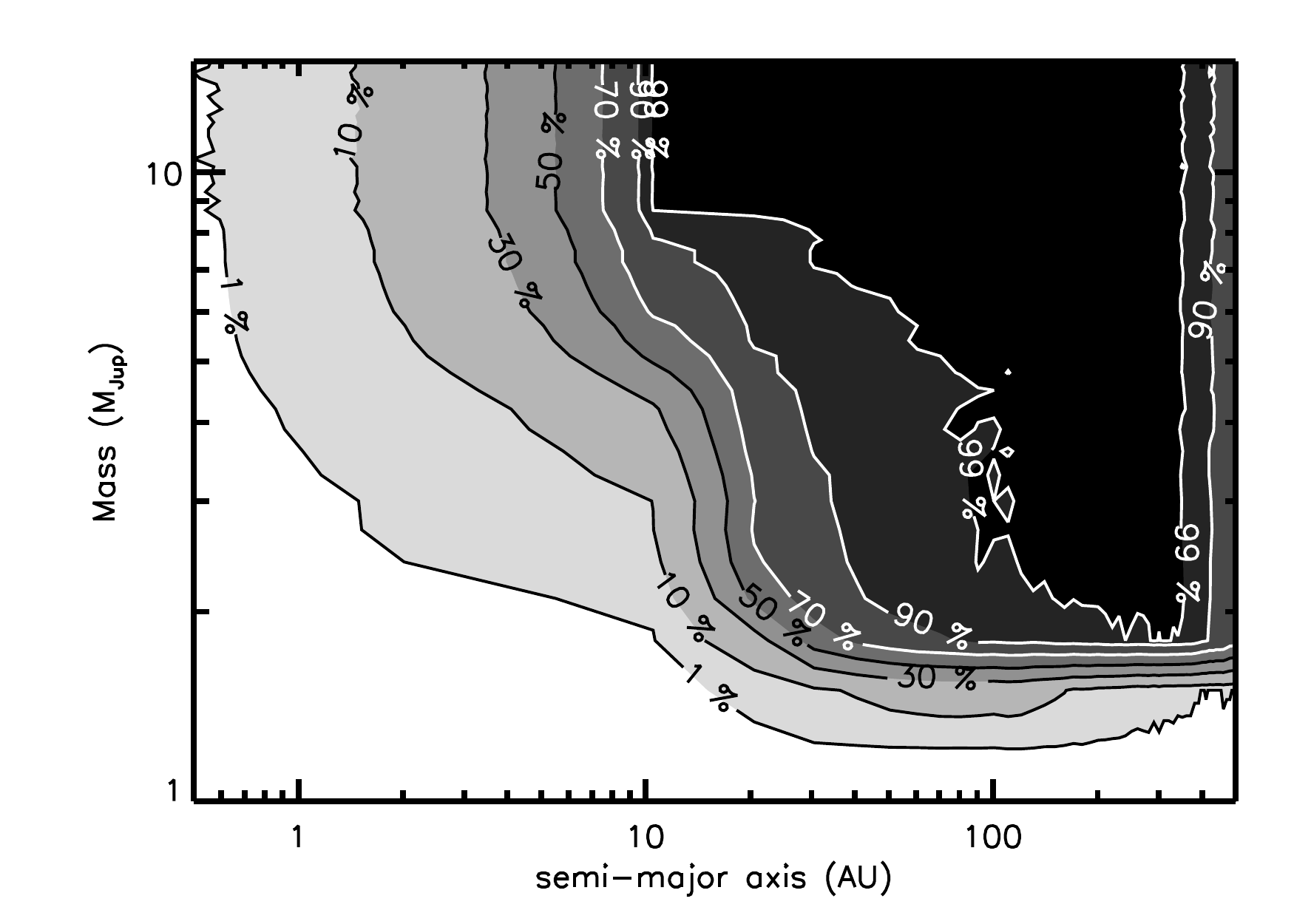}  \\ 
2M1207 &TWA25& TYC7443-1102 \\ \hline
\includegraphics[width=5.5cm]{ProbGRID_AUMic_hyb_big_2D.pdf}  &
\includegraphics[width=5.5cm]{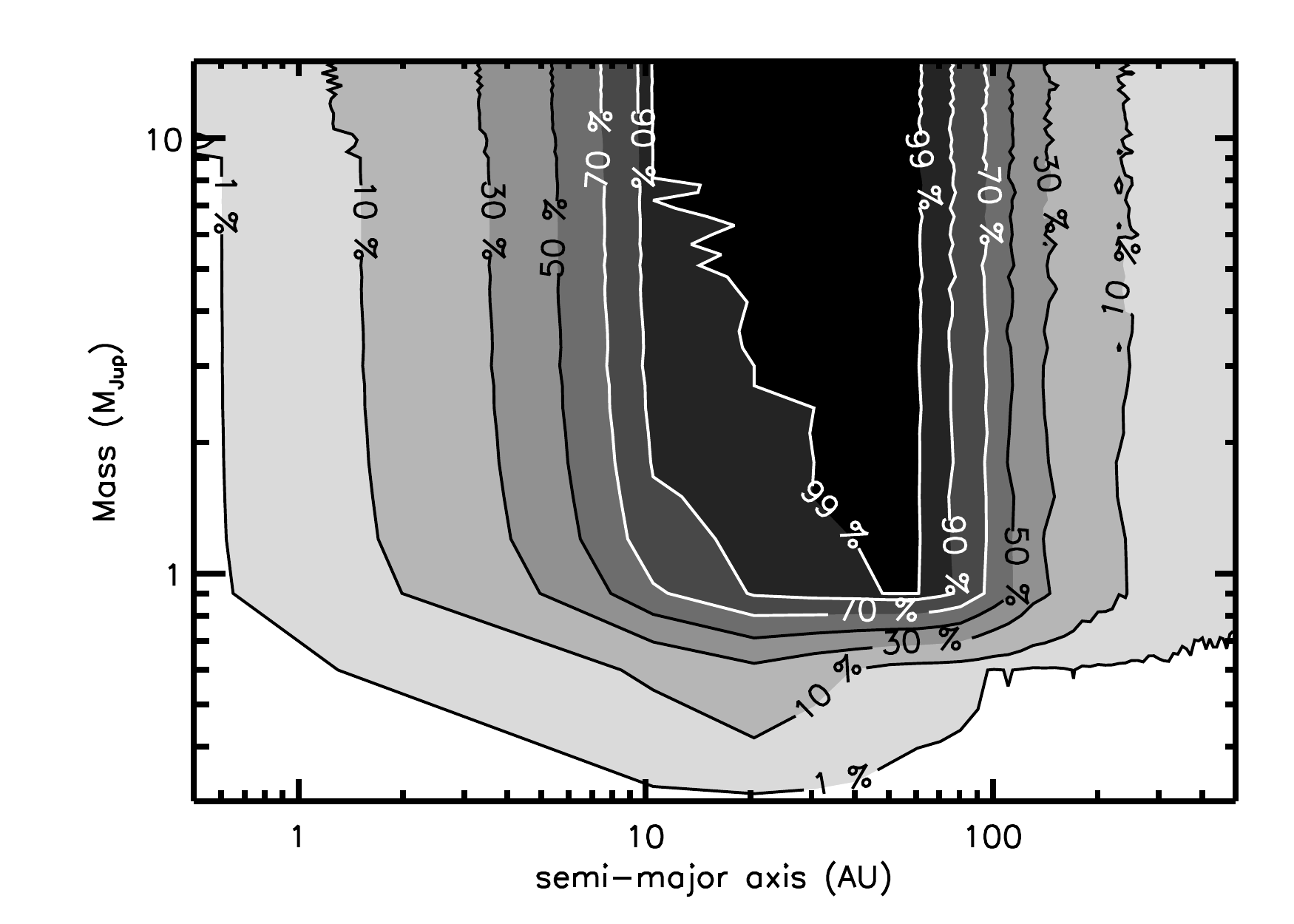} &\includegraphics[width=5.5cm]{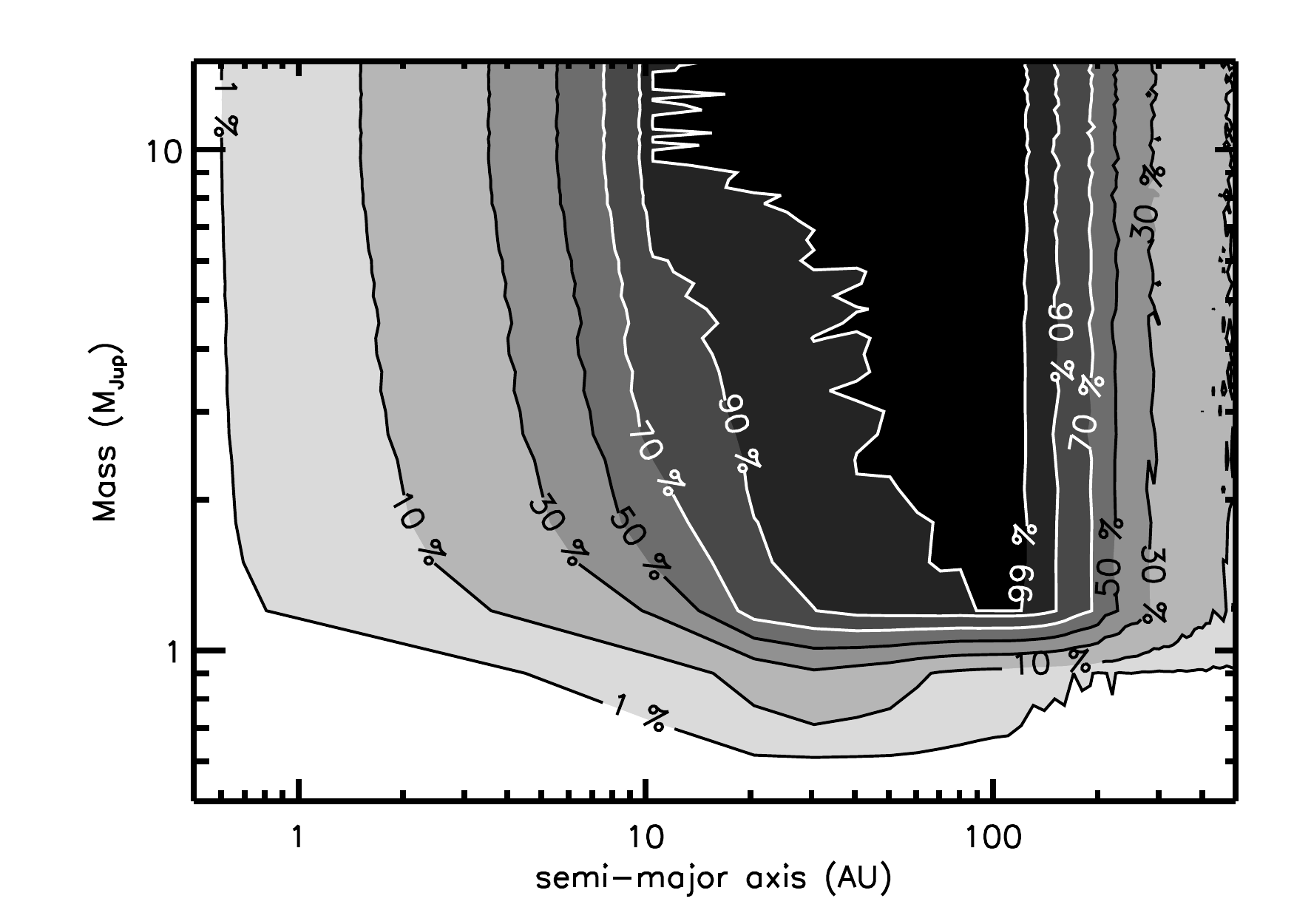}  \\ 
 AUMic/HIP102409 & WWPsA & TXPsA \\ \hline
\includegraphics[width=5.5cm]{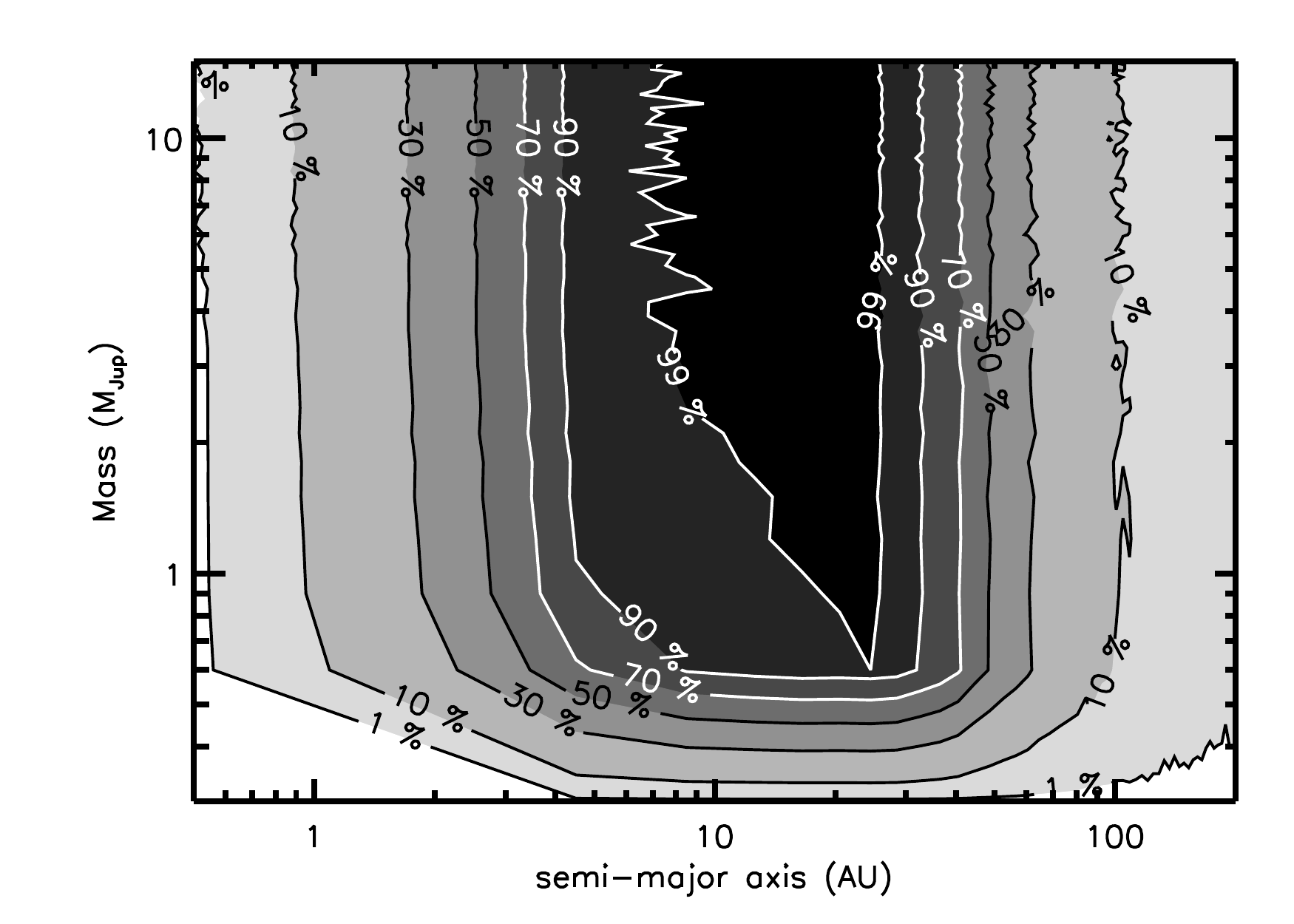}  &
\includegraphics[width=5.5cm]{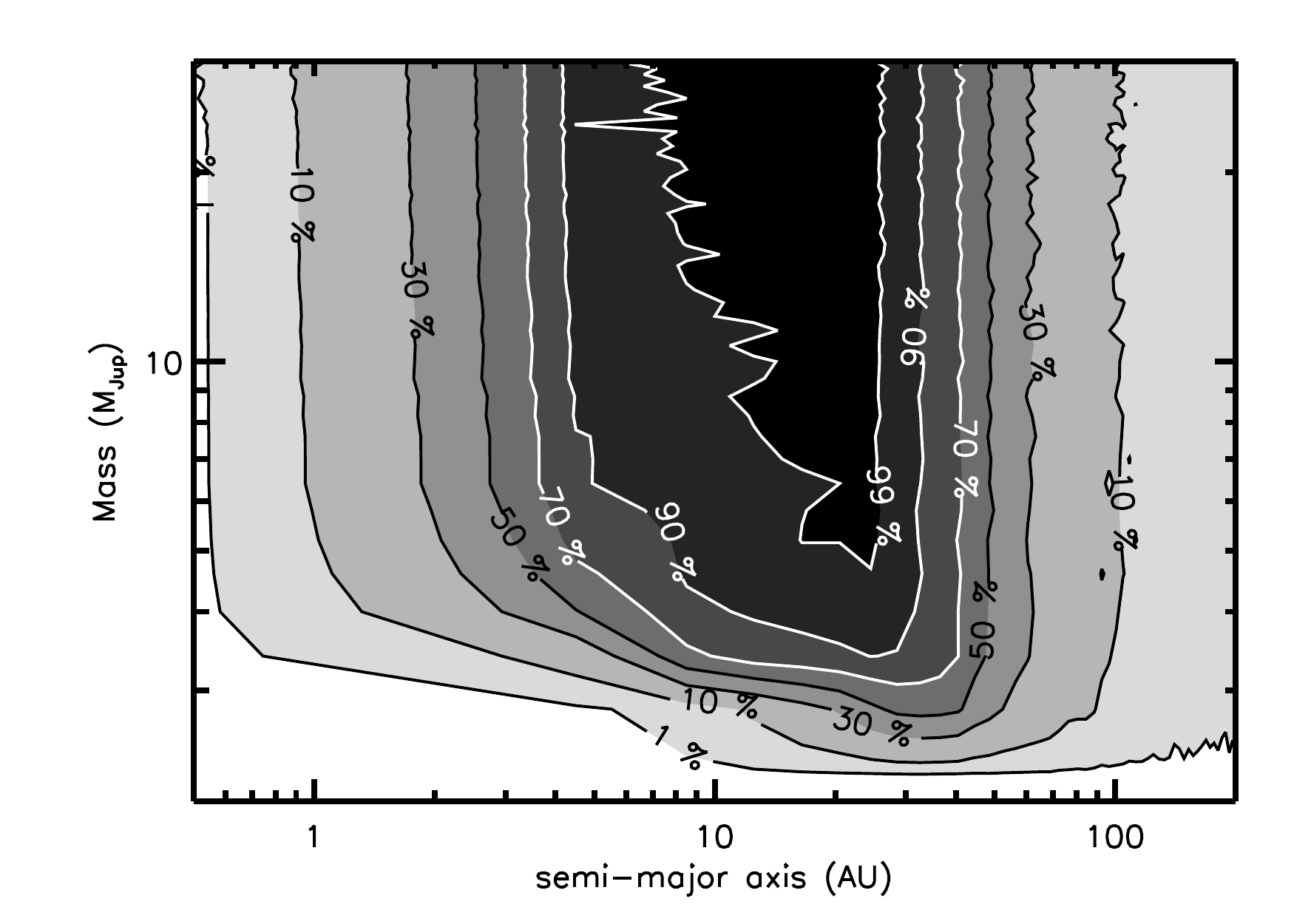}  &
\includegraphics[width=5.5cm]{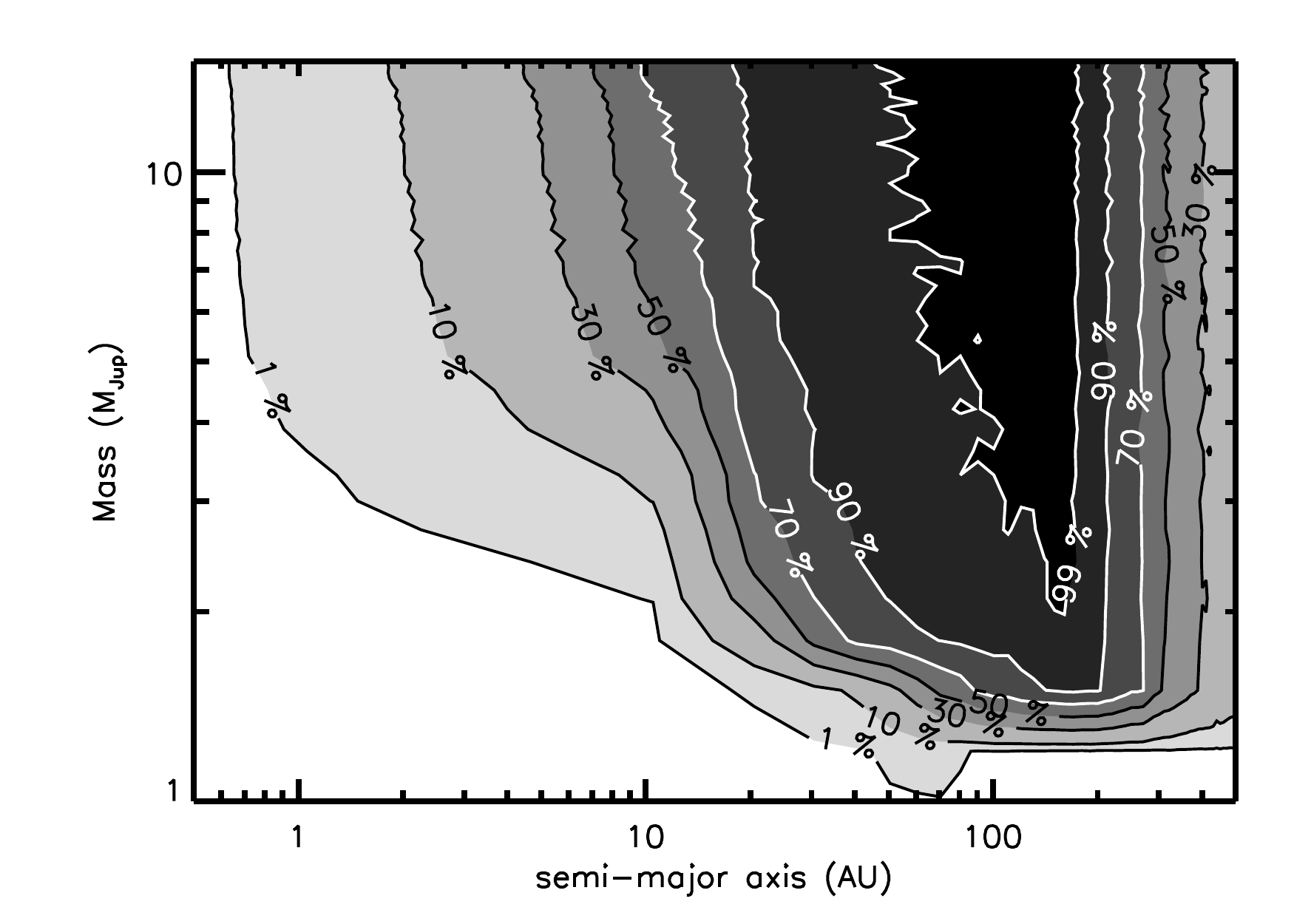} \\ 
 HIP114046 (if aged 100Myr) & HIP114046 (if aged 2000Myr)  & BD-13 6424\\ 

\end{tabular}
\end{figure*}

\begin{figure}
\includegraphics[width=9.4cm]{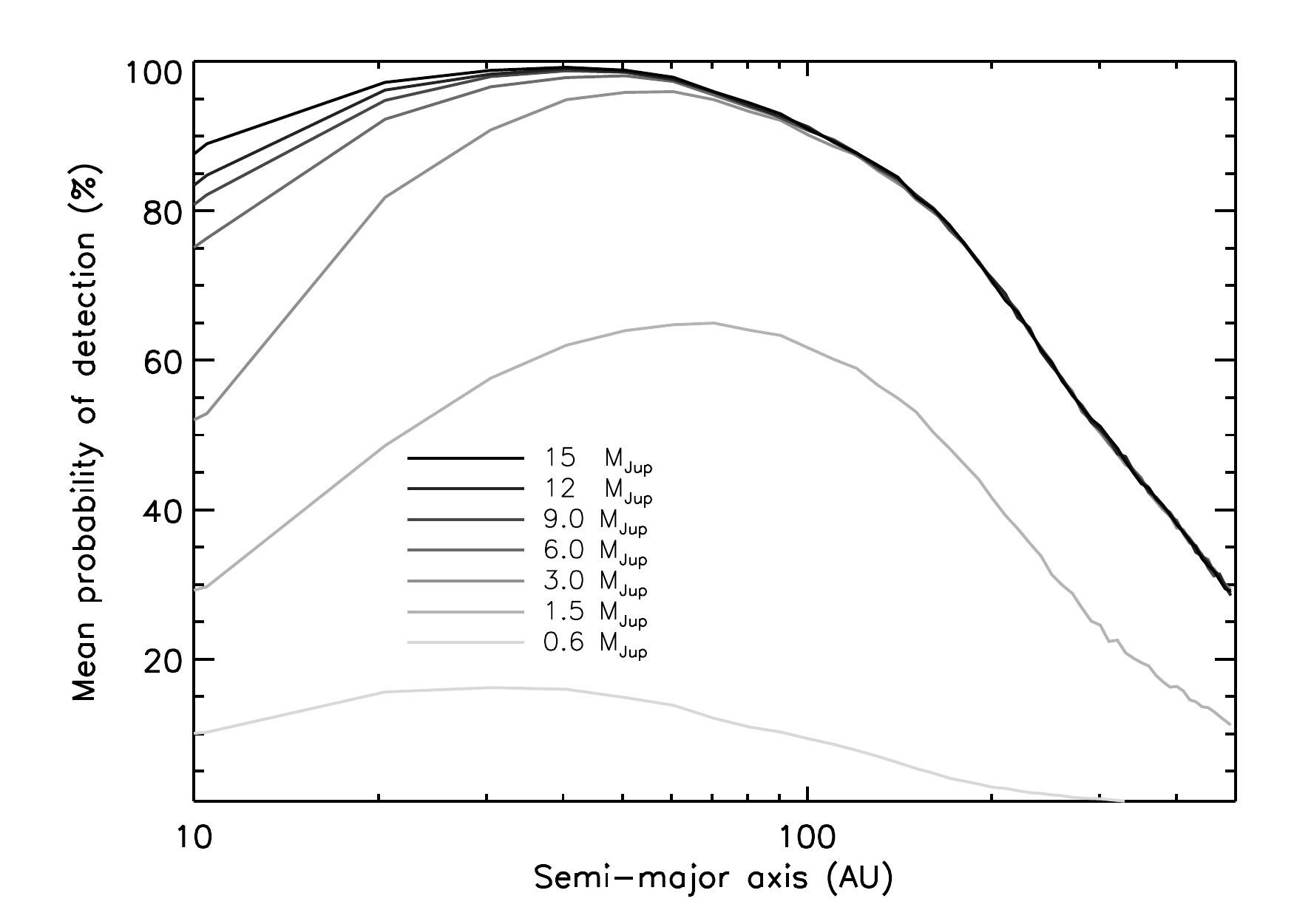}
\caption{ Summary of the detection probabilities of the survey for a range of
  companion masses, obtained by
  averaging the detection probabilities of the 14 targets with
  an accurate age determination. The semi-major axis values are therefore in real AU and not in projected AU.
\label{detlim_fullsurvey}}
\end{figure}

The resulting detection probability for each target are
showed in Fig. \ref{detlim_MESS_map}. Figure
\ref{detlim_fullsurvey} illustrates the
mass detection limits of the full survey, by averaging the 14
mass detection probabilities maps of our targets with an accurate age
measurment, and therefore excluding HIP114046 and 2M0443.  The
  decreasing detection probability for very large semi-major axis
  reflects the fact that such objects can be observed within our
  19.5$\times$19.5$\arcsec$ field of vue only on a fraction of their
  orbit and for favorable combination of excentricity and angle of sight. Using the full 52 star
sample, these limits could be used to derive constraints
on the existence of \gp around late-type stars and consequently on
planetary formation models around low-mass stars. However, the sub-sample of
16 stars we present here is too small to be derive meaningful statistics
and more observations are needed to bring it to a statistically more robust size.

\section{Conclusion}
  We presented the results of the deepest imaging survey of young M
  dwarfs to date,
  using $L'$ imaging with NACO at VLT. After developing a dedicated
  reduction and analysis pipeline, we achieved detections limits
  in average down to 1.5\mj beyond 20\,AU and up to 100-200\,AU, and 3\mj at
  10\,AU. On the closest and latest type targets we achieved detection
  limits well below the mass of Jupiter beyond 10\,AU, therefore
  actually starting to probe (on 5 objects) the mass/separation range where planets
  around M dwarfs are supposed to be be more frequent \citep[][Bonfils
  et al, submitted]{Gould.2010}, but found none.  We also probed the
  high planetary mass range (M$<$13\mj) at close separations, reaching
  planetary sensitivity at 
  5\,AU or less on 9 out of our 16 targets. In spite of these deep
  observations we found only one planetary companion,
  2M1207B, discovered by \citet{Chauvin.2004}, in this sample 
  of young M dwarfs. Unfortunately, our sample is currently too small to
  derive meaningful constraints on the existence of \gp around late
  type stars, beyond the simple statement that \gp more massive than
  1\mj are not 
  common. With the same statistical limitations, our data also
  support the ``brown dwarfs desert" hypothesis \citep{Halbwachs.2000}
 down to the lowest brown dwarfs masses, since 
  we found no brown dwarfs companions while our survey had in average more
  than 95\% de-projected sensitivity to
  brown dwarfs beyond 15\,AU and could discover a fraction of them as
  close as 5\,AU from the central star.
 Further deep observations in $L'$ of such targets are
  necessary to increase our statistical significance and to bring
  stronger constraints on formation models around low-mass stars.

\begin{acknowledgements}
We acknowledge financial support from the French Programme National de
Plan\'etologie (PNP, INSU) and from the French National Research
Agency (ANR) through the GuEPARD project grant ANR10-BLANC0504-01. P.D.
was financed by a grant from the Del Duca foundation. 
We thank David Lafreni\`ere for his precious help with the LOCI code. 
This research has made use of the VizieR catalogue access tool,
 of SIMBAD database and of Aladin, operated at CDS, Strasbourg. We
 thank the referee for his accurate comments.
\end{acknowledgements}

\bibliographystyle{aa}
\bibliography{bib}

\end{document}